\documentclass[aps,prd,amsmath,onecolumn,notitlepage,showpacs,superscriptaddress,nofootinbib,usenatbib,11pt]{revtex4-1}

\usepackage[utf8]{inputenc}
\usepackage{graphicx}
\usepackage{float}
\usepackage{booktabs}
\usepackage{amsmath}
\usepackage{fixmath}
\usepackage{amssymb}
\usepackage{hyperref}
\usepackage{comment}
\usepackage{color,xcolor}
\usepackage[normalem]{ulem}

\newcommand{\bdm}{\begin{displaymath}}
\newcommand{\edm}{\end{displaymath}}
\newcommand{\be}{\begin{equation}}
\newcommand{\ee}{\end{equation}}
\newcommand{\ba}{\begin{array}}
\newcommand{\ea}{\end{array}}
\newcommand{\pa}[1]{\left(#1\right)}
\newcommand{\paq}[1]{\left[#1\right]}
\newcommand{\ds}{\displaystyle}

\def\beq{\begin{equation}}
\def\eeq{\end{equation}}
\def\bry{\begin{eqnarray}}
\def\ery{\end{eqnarray}}

\begin{document}

\title{Cosmography with Standard Sirens from the Einstein Telescope}

\author{Josiel Mendonça Soares de Souza}
\affiliation{Departamento de F\'\i sica Te\'orica e Experimental, Universidade Federal do Rio Grande do Norte,  59072-970, Natal - RN, Brazil}
%\author[a]{Josiel Mendonça Soares de Souza}
%\affiliation[a]{Departamento de Física Teórica e Experimental, Universidade Federal do Rio Grande do Norte, Natal, RN, Brazil}
%\emailAdd{josiel.mendonca.064@ufrn.edu.br}
\author{Riccardo Sturani}
\affiliation{International Institute of Physics,\\ Universidade Federal do Rio Grande do Norte, 59078-970, Natal - RN, Brazil}
%\author[b]{Riccardo Sturani}
%\affiliation[b]{International Institute of Physics, Universidade Federal do Rio Grande do Norte, Natal, RN, Brazil}
%\emailAdd{riccardo.sturani@ufrn.br}
\author{Jailson Alcaniz}
\affiliation{Observatório Nacional, 20921-400, Rio de Janeiro - RJ, Brazil}
%\author[c]{Jailson Alcaniz}
%\affiliation[c]{Observatório Nacional, Rio de Janeiro, RJ, Brazil}
%\emailAdd{alcaniz@on.br}

\begin{abstract}
 We discuss the power of third-generation gravitational wave
  detectors to constrain cosmographic parameters
  in the case of electromagnetically bright standard sirens focusing on the specific case of the Einstein Telescope.
  We analyze the impact that the redshift source distribution, the number of detections
  and the observational error in the luminosity distance have on the inference
  of the first cosmographic parameters, and show
  that with a few hundreds detections the Hubble constant
  can be recovered at sub-percent level whereas the deceleration parameter at a few
  percent level, both with negligible bias.
\end{abstract}

\keywords{Cosmology -- Distance Scale, Gravitational Waves, Standard Sirens, Cosmological parameters.}
		
\maketitle

\section{Introduction}
\label{sec:intro}
The rising of Gravitational Wave (GW) Astronomy has opened new ways to
investigate phenomena not only in astronomy, but also in fundamental physics and
cosmology.
The large interferometric \emph{second generation} detectors LIGO \cite{TheLIGOScientific:2014jea} and
Virgo \cite{TheVirgo:2014hva} are presently in their \emph{advanced} phase,
and they will be joined in the next observation run by KAGRA \cite{KAGRA:2020tym}, 
having already confirmed 90 detections \cite{LIGOScientific:2018mvr,LIGOScientific:2020ibl,LIGOScientific:2021djp}.
In the next decade \emph{third}-generation GW detectors, like Einstein Telescope
\cite{punturo2010third} and Cosmic Explorer \cite{Reitze:2019iox},
will join the scientific quest for GWs and detections of coalescing binaries
will become a routine in astronomy, being already at the level of $\sim\, O(1)$
per week with second-generation detectors.

For cosmological applications, GWs from coalescing binaries are an invaluable
tool as they are \emph{standard sirens}
\cite{Schutz:1986gp} (see also \cite{Holz:2005df,Dalal:2006qt,Nissanke:2009kt}):
the intrinsic property of a binary system, like masses
and spins, can be inferred from a detailed analysis of the characteristic
\emph{chirp} shape of the signal,
leading to an unbiased estimation of the luminosity distance, modulo a
degeneracy with some geometric orientation angles, which can in principle
be disentangled by observation via multiple detectors, see e.g.
\cite{Vitale:2016icu} for explicit simulations.
It follows that an outstanding cosmological application of GW detections from
coalescing binaries is to measure the distance versus redshift relation,
if the redshift of the source is available.
On general grounds GW detections cannot measure the redshift of the source,
but redshift can be estimated by adding extra information, like
host galaxy identification via electromagnetic counterpart, as it happened for
the GW170817 detected jointly with the short GRB 170817A and the optical
transient AT 2017gfo \cite{LIGOScientific:2017ync},
or using galaxy catalogs either via statistical identification of host galaxies
\cite{Schutz:1986gp,DelPozzo:2011vcw,DES:2019ccw}, or by a full
cross-correlation of GW sky localization with spatial correlation functions
\cite{Oguri:2016dgk,Mukherjee:2020hyn,Diaz:2021pem}, see also \cite{Canas-Herrera:2019npr,Canas-Herrera:2021qxs,Zhang:2019loq,Jin:2020hmc,Jin:2021pcv}.

Other possibilities to fold in redshift information have been investigated
using e.g. tidal effects which can break the gravitational mass-redshift degeneracy \cite{Messenger:2011gi},
the astrophysical mass-gap \cite{Heger:2002by} in solar mass black holes,
as done in \cite{Farr:2019twy,Ezquiaga:2020tns,You:2020wju,Mastrogiovanni:2021wsd},
or the prior knowledge of source redshift distributions \cite{Ding:2018zrk,Leandro:2021qlc}.

In this work we focus on the case in which the host galaxy is identified
via an electromagnetic counterpart, so that redshift can be measured
with negligible error, while the luminosity distance is measured gravitationally,
aiming at determining the luminosity distance versus redshift relation in a
model-independent way, i.e. without relying in a specific background
cosmological model, adopting the \emph{cosmography} approach
\cite{Weinberg:1972kfs}. We thus investigate how well the time derivatives of the cosmological scale
factor can be determined using the information that electromagnetically bright
GW standard sirens will reveal with third-generation detectors, expected
to be operating in the next decade.

The paper is organized as follows: in Sec.~\ref{sec:cosmogr} we briefly review the cosmographic
expansion whereas the simulated data used in the statistical analysis are detailed in Sec.~\ref{sec:mock}.
The main results of the paper are presented and discussed in Sec.~\ref{sec:results}. Sec.~\ref{sec:concl} presents the conclusions that can be drawn from our analysis.

\section{The cosmographic expansion}
\label{sec:cosmogr}
Let us consider the flat Friedmann-Lema\^\i tre-Robertson-Walker metric  ($c = 1$)
\begin{equation}
  ds^2 = -dt^2+a^2(t)\left[dr^2+r^2(d\theta^2+\sin^2\theta d\phi^2)\right]\,,
\end{equation}
where $a(t)$ is the scale factor. One can Taylor-expand its time dependence
\begin{eqnarray}
  \label{eq:aoft}
  a(t) = a_0\left\lbrace1 + H_0(t-t_0) - \frac{1}{2}q_0[H_0(t-t_0)]^2 	+ \frac{1}{6}j_0[H_0(t-t_0)]^3 \right. \nonumber\\
  \left.+ \frac{1}{24}s_0[H_0(t-t_0)]^4   + \frac{1}{120}l_0[H_0(t-t_0)]^5 + O(6)\right\rbrace\,,
\end{eqnarray}
where $H_0$ is the Hubble constant, $a_0\equiv a(t_0)$ being $t_0$ the present time, and the remaining constants
appearing in the expansion have been conveniently expressed as
\be
\ba{rclrcl}
\ds q_0 &=&\ds -\frac{1}{H_0^2}\left(\frac{1}{a}\frac{d^2a}{dt^2}\right)\bigg\vert_{t=t_0}\,,&\ds \quad j_0 &=&\ds \frac{1}{H_0^3}\left(\frac{1}{a}\frac{d^3a}{dt^3}\right)\bigg\vert_{t=t_0},\\
\ds s_0 &=&\ds \frac{1}{H_0^4}\left(\frac{1}{a}\frac{d^4a}{dt^4}\right)\bigg\vert_{t=t_0}\,,&  
\ds \quad l_0 &=&\ds \frac{1}{H_0^5}\left(\frac{1}{a}\frac{d^5a}{dt^5}\right)\bigg\vert_{t=t_0}\,,
\ea
\ee
where $H\equiv a^{-1}da/dt$.
The parameters $q_0,j_0,s_0,l_0$ are called respectively \textit{deceleration},
\textit{jerk}, \textit{snap} and \textit{lerk} parameters \cite{visser2004jerk, Cattoen:2007sk, aviles2012cosmography,Lazkoz:2013ija}. From these  equations we can also obtain \cite{zhang2017we}
\begin{eqnarray}
  \label{eq:Hofz}
  H(z) = H_0\left\lbrace 1 + (1+q_0)z + \frac{1}{2}(j_0-q_0^2)z^2 + \frac{1}{6}[3q_0^2+3q_0^3-4q_0j_0-3j_0-s_0]z^3 + \right. \nonumber \\
  + \left. \frac{1}{24}(-12q_0^2-24q_0^3-15q_0^4 + 32q_0j_0 + 25q_0^2j_0 + 7q_0s_0 + 12j_0 - 4j_0^2 + 8s_0 + l_0)z^4 + O(z^5)\right\rbrace\;,
\end{eqnarray}
where $1+z\equiv 1/a(t)$. To make contact with observations it is useful to define the {luminosity distance} expression %in terms of the redshift
\begin{equation}
  \label{eq:dl}
  d_L(z)=\pa{1+z}\int_0^z\frac{dz'}{H(z')}\,,
\end{equation}
which, in terms of the cosmographic parameters  Eq.~(\ref{eq:Hofz}), can be expressed as
%\cite{aviles2012cosmography}
\begin{eqnarray}
  \label{eq:dlvsz}
  d_L(z) & = & \frac{1}{H_0} \Bigl[ z + z^2 \Bigl(\frac{1}{2} - \frac{q_0}{2} \Bigr) +
    z^3 \Bigl(-\frac{1}{6} -\frac{j_0}{6} + \frac{q_0}{6} + \frac{q_0^2}{2} \Bigr) \nonumber \\
    && +z^4 \Bigl( \frac{1}{12} + \frac{5 j_0}{24} - \frac{q_0}{12} + \frac{5 j_0 q_0}{12} -
    \frac{5 q_0^2}{8} - \frac{5 q_0^3}{8} + \frac{s_0}{24} \Bigr) + \nonumber\\
    && +z^5 \Bigl( -\frac{1}{20} - \frac{9 j_0}{40} + \frac{j_0^2}{12} - \frac{l_0}{120} +
    \frac{q_0}{20} - \frac{11 j_0 q_0}{12} + \frac{27 q_0^2}{40} - \frac{7 j_0 q_0^2}{8} + \frac{11 q_0^3}{8} +
    \frac{7 q_0^4}{8} - \frac{11 s_0}{120} - \frac{q_0 s_0}{8} \Bigr)\nonumber\\
    && +z^6 \Bigl( \frac{1}{30} + \frac{7 j_0}{30} - \frac{19 j_0^2}{72} + \frac{19 l_0}{720} +
    \frac{m_0}{720} - \frac{q_0}{30} + \frac{13 j_0 q_0}{9} - \frac{7 j_0^2 q_0}{18} + \frac{7 l_0 q_0}{240}
    - \frac{7 q_0^2}{10} + \frac{133 j_0 q_0^2}{48}\nonumber\\
    &&  - \frac{13 q_0^3}{6} + \frac{7 j_0 q_0^3}{4} - \frac{133 q_0^4}{48} - \frac{21 q_0^5}{16} + \frac{13 s_0}{90}
    - \frac{7 j_0 s_0}{144} + \frac{19 q_0 s_0}{48} + \frac{7 q_0^2 s_0}{24} \Bigr) +O(z^7)\Bigr].
\end{eqnarray}
Observation of a set of $\{d_{Li}\}$ and corresponding $\{z_i\}$ data points
will enable the determination of the cosmographic parameters, whose accuracy
will depend on the quality and range of available data, as we will discuss in the next Section.
Note that any finite order truncation of the series (\ref{eq:dlvsz}) is badly behaved for $z>1$.
However, as it will be shown, for the range of redshift of interest for the
present work, we can safely work with (a truncation of) Eq.~(\ref{eq:dlvsz}).

\section{Simulated data}
\label{sec:mock}

To test the accuracy of the cosmographic expansion applied to GW standard sirens,
we set the fiducial model to be the flat $\Lambda$CDM cosmology, with the matter density parameter $\Omega_M=0.311$ and
$H_0=67.66$ km/sec/Mpc, in agreement with the current Cosmic Microwave Background analysis of \cite{Planck:2018vyg}.
Note that the
present controversy on the value of $H_0$ -- with low-redshift determinations in contrast with early-Universe ones by more than 4$\sigma$ level~\cite{2018ApJ...855..136R,Birrer:2018vtm,Verde:2019ivm,refId0} --
does not affect our analysis, which aims at quantifying the
uncertainties on the cosmographic parameters expected from upcoming joint GW and electromagnetic data.\footnote{See also \cite{Dainotti:2021pqg} for an analysis
of the variation of $H_0$ taking into accoun SNe data from different redshift ranges.}

The cosmographic parameters (up to sixth order) for the $\Lambda$CDM model are written as:
\be
  \ba{rcl}
  \ds q_0 &=&\ds \frac{3}{2}\Omega_M-1\,,\\
  \ds j_0 &=&\ds 1\,,\\
  \ds s_0 &=&\ds 1-\frac{9}{2}\Omega_M\,,\\
  \ds l_0 &=&\ds 1 + 3\Omega_M-\frac{27}{2}\Omega_M^2\,,\\
  \ds m_0 &=&\ds 1 - \frac{27}{2}\Omega_M^2 - 81\Omega_M^2 - \frac{81}{2}\Omega_M^3\,,
  \ea
\ee
whose numerical values for our fiducial model are summarized in Tab.~\ref{tab:cosmopars}.

\begin{table}[ht]
  \centering 
  \begin{tabular}{ccccc} \hline \hline
    $q_0$ &	$j_0$ &	$s_0$ &	$l_0$ & $m_0$\\
    $\quad-0.533\quad$ & $\quad 1\quad$ & $\quad-0.4\quad$ & $\quad 0.627\quad$ & $\quad -9.365\quad$ \\
    \hline \hline		
  \end{tabular}
  \caption{Cosmographic parameters in a flat $\Lambda$CDM cosmology with $\Omega_M=0.311$.}%, $\Omega_\Lambda=1-\Omega_M$. The contribution to the right hand side of eq.~(\ref{eq:G00}) from radiation and spatial curvature are assumed to be vanishing.}
\label{tab:cosmopars}
\end{table}

\begin{figure}[t]
  \begin{center}
    \includegraphics[width=.49\linewidth]{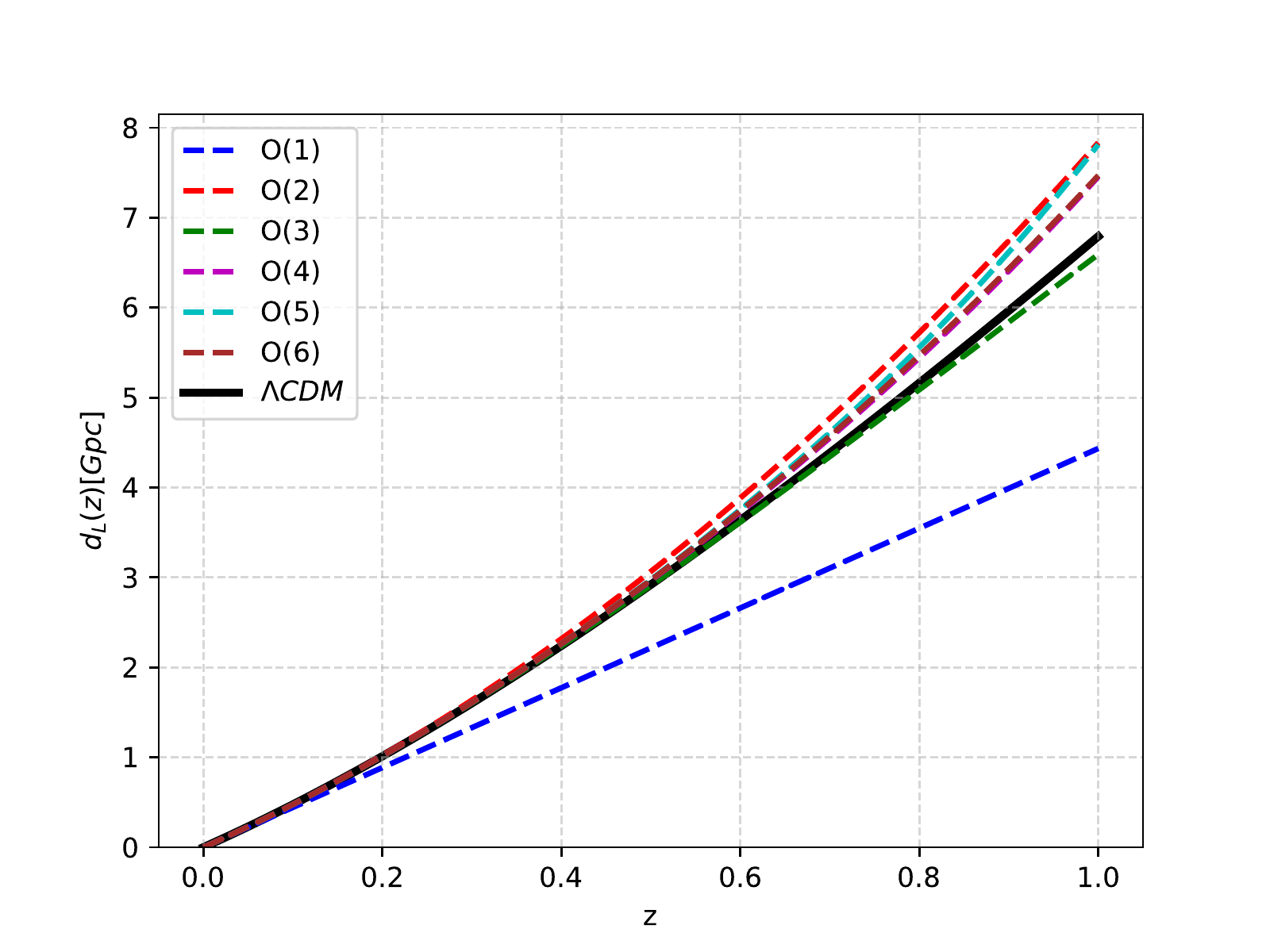}
    \includegraphics[width=.49\linewidth]{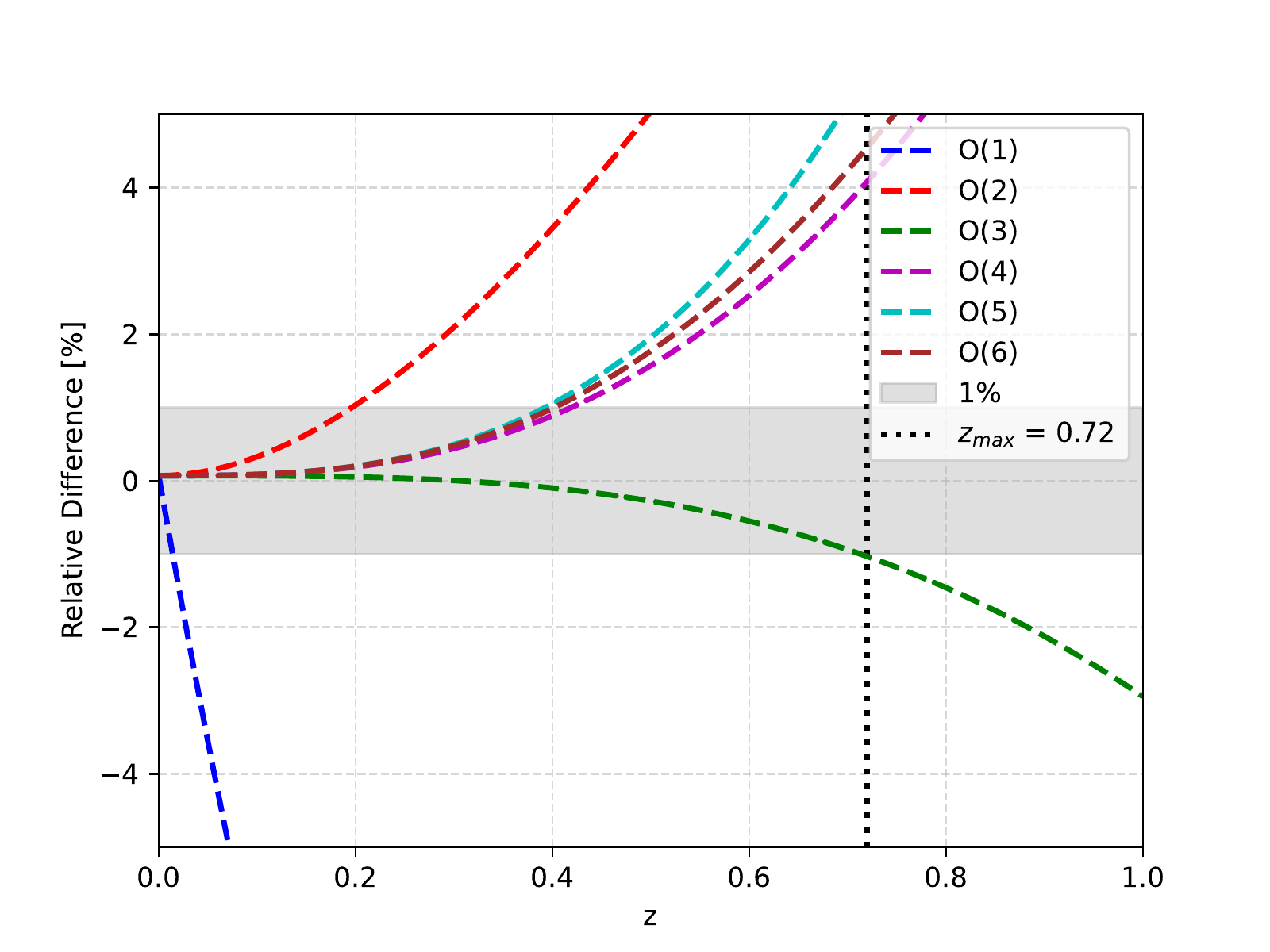}
    \caption{{\it{Left}}) Luminosity distance $d_L$ for different orders of the cosmographic series -- Eq. (\ref{eq:dlvsz}).
      The black curve represents the fiducial $\Lambda$CDM model adopted in the analysis.
      {\it{Right}}) Relative difference between the cosmographic series and the fiducial cosmology.
      The shadowed horizontal region stands for a difference of 1\%, which is reached  at $z_{cut} = 0.72$ (vertical dotted line) by the third order series.}
    \label{fig:dLz}
  \end{center}
\end{figure}

As mentioned earlier, the $d_L$ expansion in $z$ converges for $z<1$ \cite{Cattoen:2007sk}. However, to have a faithful
reconstruction of the underlying model with only a handful of terms one needs to restrict the redshift to lower values.
In Fig.~\ref{fig:dLz} (left panel) we show the behavior of $d_L$ as a function of the redshift for different orders
of the cosmographic series given in Eq.(\ref{eq:dlvsz}), with values of the
cosmographic parameters reported in Tab.~\ref{tab:cosmopars}.
For comparison, our fiducial model is also shown (black curve).
The right panel shows the relative difference between the cosmographic series and the fiducial cosmology. As it can be seen,
the series truncated at third order deviates less than $1\%$ from the fiducial model in the redshift interval $0<z<0.72$,
which in turn defines the redshift interval
adopted in this work to assess the precision with which one can recover
the first three cosmographic parameters. %: $H_0$, $q_0$, and $j_0$.
Note that the cosmographic parameter values reported in Tab.~\ref{tab:cosmopars}
are the ones which reproduce exactly the Taylor expansion of the $\Lambda$CDM
model with $\Omega_M=0.311$, hence it is expected that truncating the
$z$-expansion at any finite order, the best fit values of the cosmographic
parameters will be different from these ones.\footnote{We verified that for
  realistic values of $h$ in the interval [0.65-0.75] and $\Omega_m\sim 0.3$,
  the value of $z_{cut}$ does not change significantly.
  A more general approach using non-parametric methods
  (e.g.~Gaussian Process \cite{Seikel:2012uu,Belgacem:2019zzu,Canas-Herrera:2021qxs}) can be employed if alternative
  cosmologies are considered.}

\begin{figure}[t]
  \begin{center}
    \includegraphics[width=.49\linewidth]{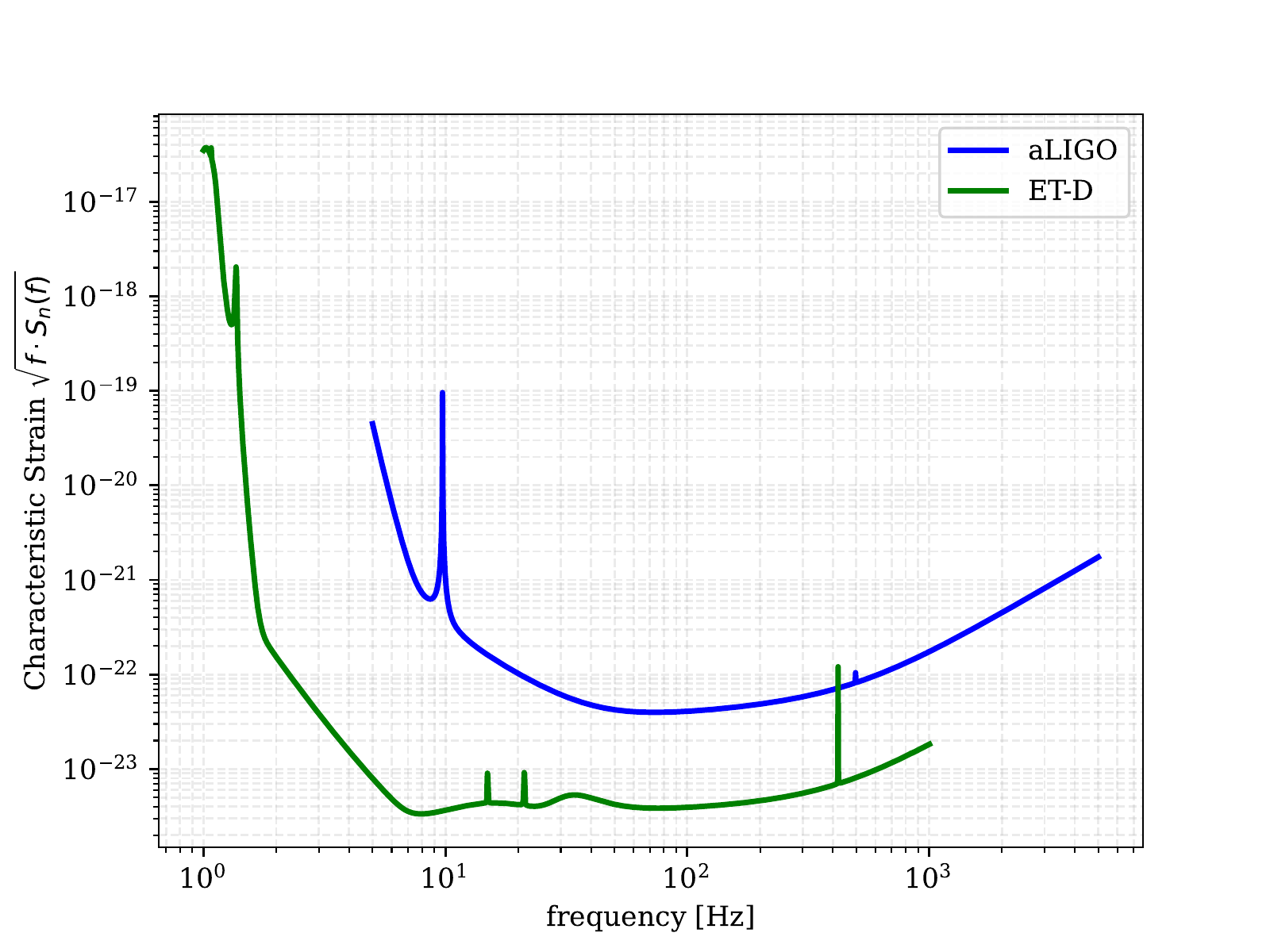}
    \includegraphics[width=.49\linewidth]{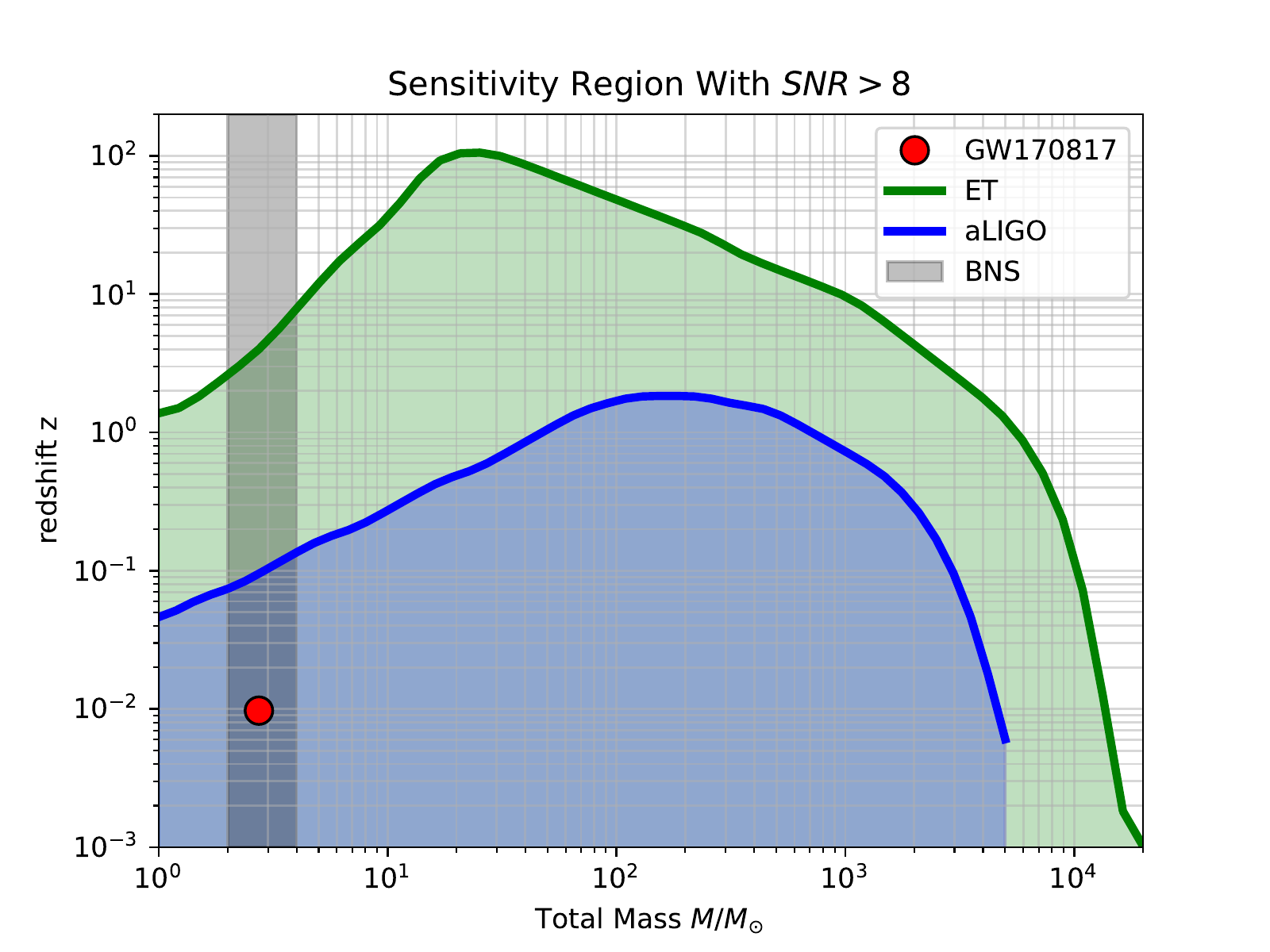}
    \caption{{\it{Left}}) Dimensionless strain sensitivity for Advanced LIGO, from ``Zero Detuning High Power'' spectral noise density implemented in LALSuite \cite{lalsuite},
      and Einstein Telescope, adopting the noise spectral density ``D'' from \cite{Hall:2019xmm}. {\it{Right}}) Sensitivity region, defined as the distance at
      which the $SNR$ of Eq.~(\ref{eq:snr}) is larger than the conventional threshdold
      value $8$, for spin-less equal mass systems with optimal orientation with
      respect to the detector (only the GW dominant mode has been used). The dashed area highlights total mass values for which an electromagnetic counterpart is expected.}
    \label{fig:noise_reach}
  \end{center}
\end{figure}

\subsection{Source Distributions}
\label{ssec:s_dists}
For the simulation of third generation detector observations we focus on
the Einstein Telescope \cite{punturo2010third}, whose noise spectral density is report in Fig.~(\ref{fig:noise_reach}) together with the advanced LIGO one.

The signal-to-noise ratio $SNR$ of a real-time GW signal $h(t)$ is defined in terms
of its Fourier-transform $\tilde h(f)$ via\footnote{Here, 
  $h(t)$ is the scalar time series representing the projection onto the detector
  of the two GW polarizations.}
\be
\label{eq:snr}
SNR^2\equiv 4\int_0^\infty \frac{|\tilde h(f)|^2}{S_n(f)}\, df\,,
\ee
where the noise spectral density $S_n(f)$ is defined in terms of the
detector noise $\tilde n(f)$ averaged over many realizations
\be
\langle\tilde n(f)\tilde n^*(f')\rangle=S_n(f)\delta(f-f')\,.
\ee

\begin{figure}
  \begin{center}
  \includegraphics[width=.45\linewidth]{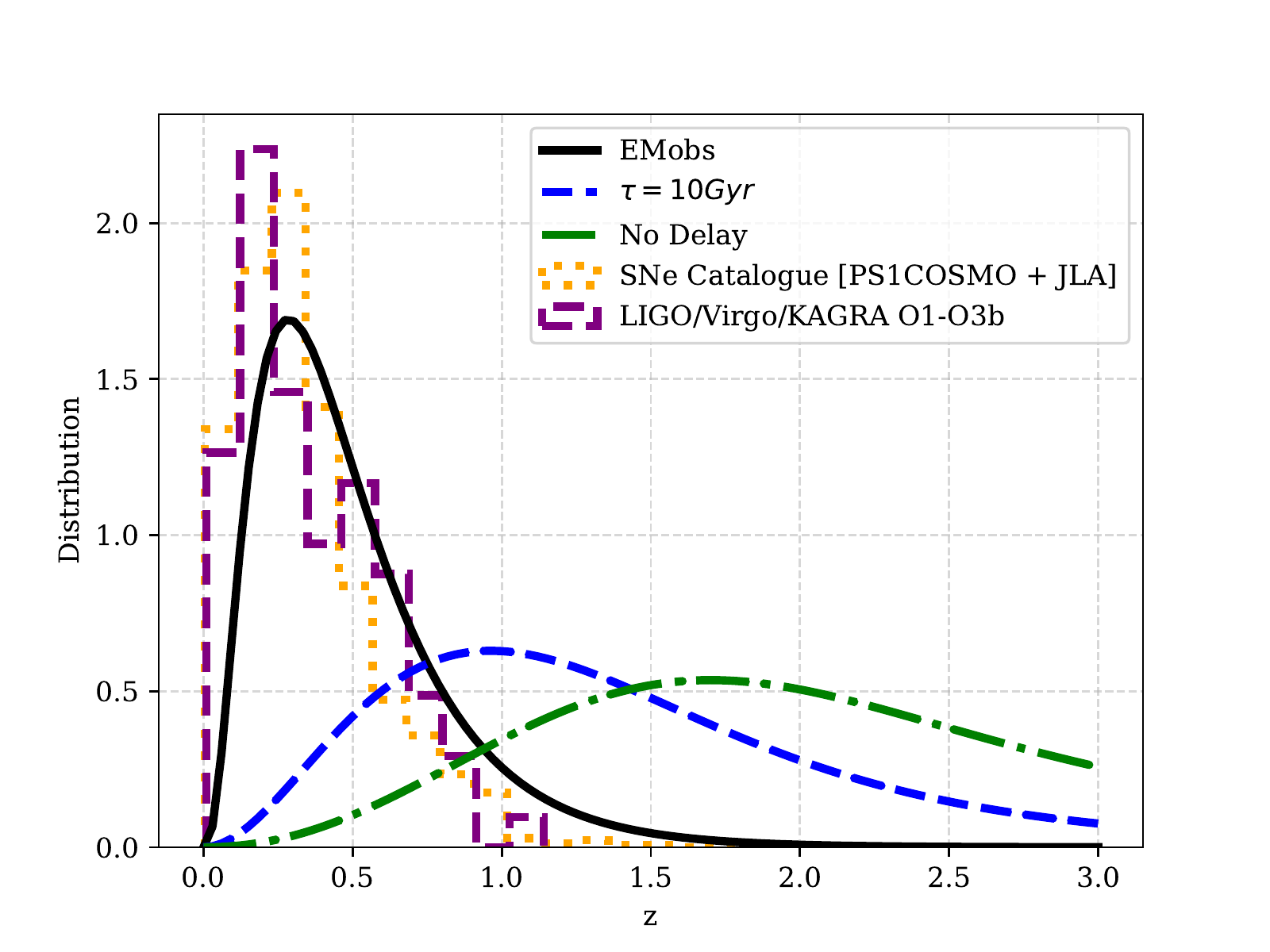}
  \includegraphics[width=.45\linewidth]{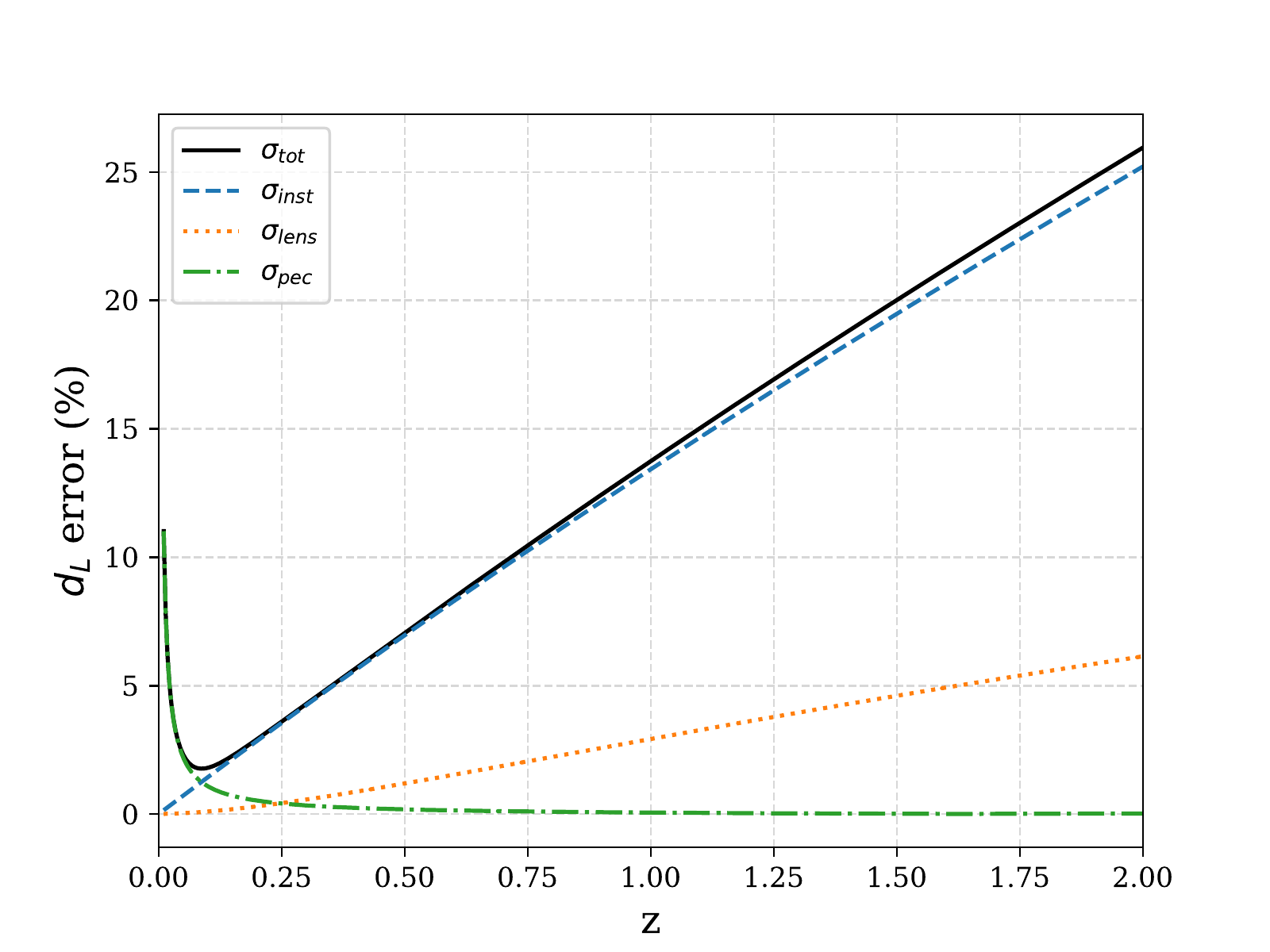}
  \caption{{\it{Left}}) Expected normalized distributions of EM bright GW events detected in coincidence
    by the Einstein Telescope and future EM observatory Theseus Eq.~(\ref{eq:EMobs}) \cite{Belgacem:2019tbw} or dictated by (delayed) star formation rate (Eq.~\ref{eq:mr}).
    For comparison, the normalized distributions of presently confirmed
    events by the LIGO/Virgo/KAGRA collaboration and of the SuperNovae catalog
    events \cite{betoule2014improved,2018ApJ...859..101S,2018ApJ...857...51J}
    are also reported. {\it{Right}}) Error budget of luminosity distance gravitational measure -- see Eqs.~(\ref{eq:dL_err}) and (\ref{eq:dL_errs}).}
  \label{fig:dists}
  \end{center}
\end{figure}

The redshift distribution of coalescing events is largely unknown, hence to
make an analysis as comprehensive as possible we follow \cite{deSouza:2019ype}
and produce and analyze mock data corresponding to
three astrophysical source distributions to check how they affect
the parameter estimation. We thus consider:

\begin{enumerate}
\item Detected merger distribution per comoving volume following the star formation rate given in \cite{madau2014cosmic}:
  \begin{equation}
    \label{eq:sfr}
    \psi(z)\propto\frac{\pa{1+z}^{2.7}}{1+\pa{\frac{1+z}{2.9}}^{5.6}}\,,
  \end{equation}
\item {Merger distribution obtained with a stochastic delay with respect to
  the star formation rate, the delay being Poisson-distributed with average
  $\tau=10$ Gyr, as shown in \citep{vitale2017parameter}, with
  a merger rate per unit comoving volume ${\cal R}_m$ given by (modulo normalization)}
  \be
    \label{eq:den_mr}
  {\cal R}_m(z_m, \tau)\propto\int_{z_m}^\infty dz_{f} \frac{dt}{dz_{f}}\psi(z_{f})
  \exp\paq{-\frac{t(z_{f})-t(z_m)}\tau}\,.
  \ee
  Note that in both this and the previous case, the detected merger rate $R_m$
  is related to the density merger rate of (\ref{eq:den_mr}) via
  \be
  \label{eq:mr}
  \ba{rcl}
  \ds R_m(z, \tau)&\equiv&\ds\frac{d^2N_m}{dt_odz}=\frac 1{1+z}\frac{dV_c}{dz}{\cal R}_m(z, \tau)\\
  &=&\ds \frac{4\pi d_L^2(z)}{(1+z)^3H(z)}{\cal R}_m(z, \tau)\,.
  \ea
  \ee
\item A distribution considering that only close enough
  events will have a detectable EM counterpart. With a dedicated EM
  detector like THESEUS \cite{amati2018theseus,Belgacem:2019tbw} it can be
  represented as shown in Fig.~(\ref{fig:dists}) (``EMobs'' line) \cite{deSouza:2019ype}
  \begin{equation}
    \label{eq:EMobs}
    R_{EMobs} \propto \frac{z^3}{1+\exp(10.6\, z^{0.6})}\,.
  \end{equation}
\end{enumerate}

\subsection{Luminosity distance uncertainty}
\label{ssec:uncert}
\begin{figure}[t]
  \centering
  \includegraphics[width=.5\linewidth]{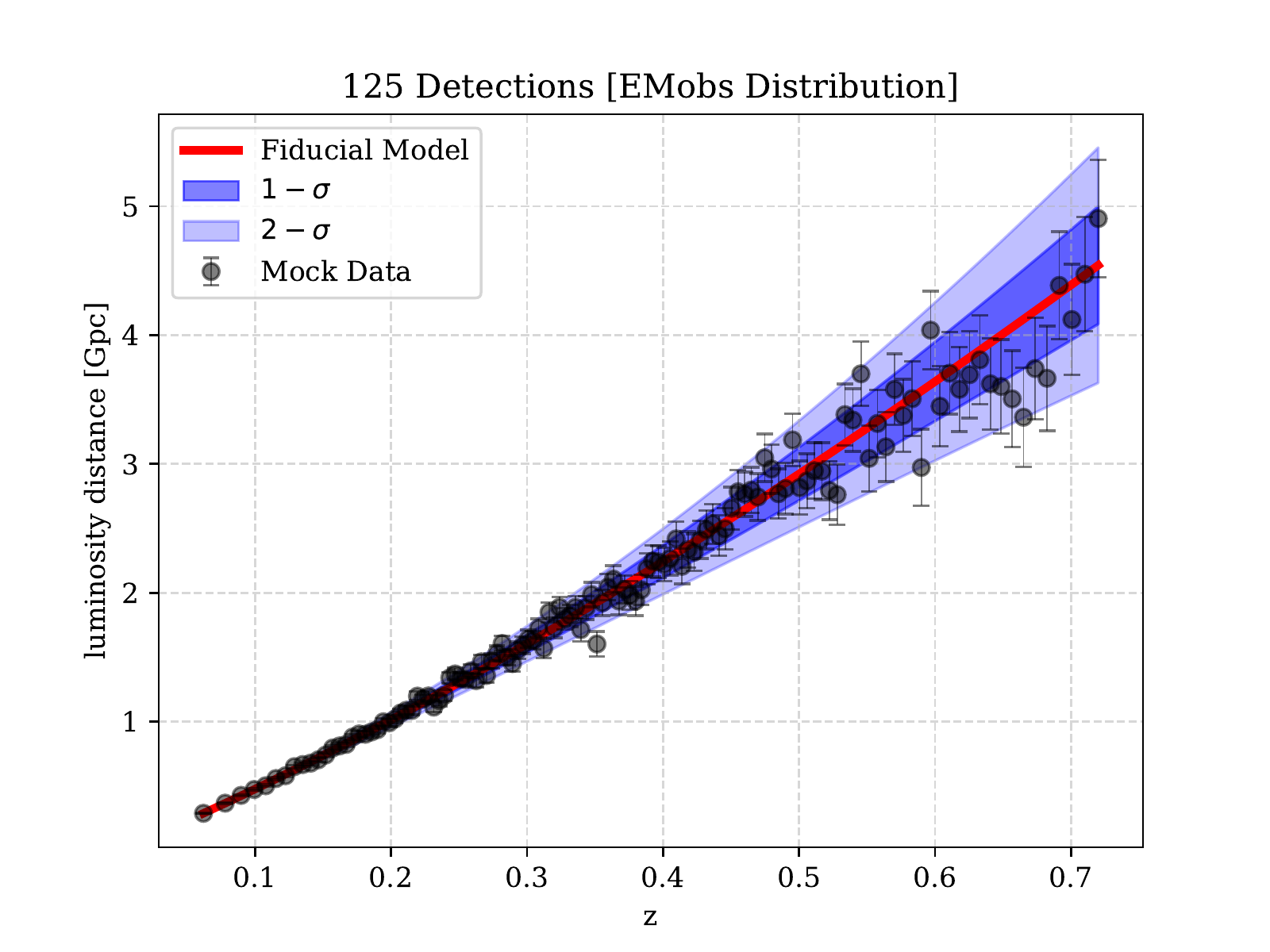}
  \caption{Mock data using the EMobs distribution Eq.~(\ref{eq:EMobs}). The fiducial $\Lambda$CDM model is denoted by the red line, blue bands denote
    1- and 2-$\sigma$ confidence regions.}
\label{fig:dist_error}
\end{figure}

Having fixed the source redshift distribution, we now discuss the uncertainties in the luminosity distance measurements by the third generation gravitational wave detector \textit{Einstein Telescope}.
The main contributions are given by \cite{zhao2011determination,Gordon:2007zw}:
\begin{equation}
  \label{eq:dL_err}
  \frac{\Delta d_L(z)}{d_L(z)} =\ds \left[
    \left(\frac{\Delta d_L(z)}{d_L(z)}\right)_{ET}^2+
    \left(\frac{\Delta d_L(z)}{d_L(z)}\right)_{Lensing}^2+
    \left(\frac{\Delta d_L(z)}{d_L(z)}\right)_{v_{pec}}^2\right]^{1/2}\,,
\end{equation}
with
\begin{equation}
  \label{eq:dL_errs}
  \begin{array}{lcl}
    \ds\left(\frac{\Delta d_L(z)}{d_L(z)}\right)_{ET} &\approx&\ds
    0.1449z - 0.0118z^2 + 0.0012z^3 \quad  ,\\
    \ds\left(\frac{\Delta d_L(z)}{d_L(z)}\right)_{Lensing} &\approx&\ds
    0.066 \paq{4 \pa{1-(1+z)^{-1/4}}}^{1.8} \quad ,\\
    \ds\left(\frac{\Delta d_L(z)}{d_L(z)}\right)_{v_{pec}} &\approx&\ds \bigg| 1 - \frac{(1+z)^2}{H(z)d_L(z)} \bigg|\frac{\sigma_v}{c}, \quad \sigma_v=331\ {\rm{km/s}}\,,
  \end{array}
\end{equation}
where the error budget has been divided respectively into instrumental,
lensing and peculiar velocity contributions.
As the systematic errors due peculiar velocities of galaxies are important only
for $z\lesssim 0.1$, they do not play an important role in this work.
In Fig.~(\ref{fig:dist_error}), we show the simulated data of $d_L(z)$ using the EMobs distribution given by Eq.~(\ref{eq:EMobs}). The red curve represents the fiducial model used in the simulation whereas the light blue and dark blue regions stand respectively
  for 1- and 2-$\sigma$ error band, being $\sigma$ the quantity $\Delta d_L$, given the error
  budget (Eq.~\ref{eq:dL_err}).

\subsection{Parameter Estimation}
\label{ssec:Bayes}
We adopt a Bayesian framework to estimate the parameters
$\{\theta_i\}=\{H_0,q_0,j_0\}$ given the model $\cal M$ represented by the
cosmographic expansion (\ref{eq:dlvsz}) truncated at third order. In this
context the combined probability distribution functions of parameters $\theta_i$
are given by
\begin{equation}
  P(\lbrace \theta_i\rbrace|d,\ \mathcal{M}) = \frac{P(d|\lbrace \theta_i\rbrace,\ \mathcal{M})P(\lbrace \theta_i\rbrace|,\ \mathcal{M})}{P(d)}\,,
\end{equation}
where the likelihood $P(d|\lbrace \theta_i\rbrace,\ \mathcal{M})$ is given by:
\begin{equation}
  P(d|\lbrace \theta_i\rbrace,\ \mathcal{M}) = \exp\left[-\frac{1}{2}\sum_{i=1}^N \frac{|d_i - f_i(\lbrace \theta_i\rbrace,z_i)|}{\sigma_i^2}\right]\,,
\end{equation}
being $N$ the number of detections, $d_i$ the data with error $\sigma_i$ and
$f_i(\lbrace \theta_i\rbrace,z_i)$ is the parameter-dependent model function (third order truncation of Eq.~\ref{eq:dlvsz}) and
the priors on $H_0$, $q_0$, and $j_0$ are taken to be flat in the region
respectively $[54,81]$km/sec/Mpc, $[-1,0]$, $[-1,3]$.
Note that we have neglected the error on the redshift, consistently with the
assumption of an EM counterpart for the GW signal, that allows determination of the host galaxy determination.
To perform our analysis we use the \texttt{Bilby} software \cite{Ashton:2018jfp} with the \texttt{Nestle} sampler, \cite{Mukherjee:2005wg} which implements the nestled sampling algorithm \cite{Skilling:2006gxv}, with 500 live points. 

In the next section we estimate the errors and the biases in the cosmographic parameters $H_0$, $q_0$ and $j_0$ for the
three source distributions introduced in Subsec.~\ref{ssec:s_dists},
and discuss how such estimates change as data at different redshift ranges are considered.

\section{Results}
\label{sec:results}

As mentioned earlier, we restrict the redshift range to $0<z<0.72$,
which corresponds to the region where the third order cosmographic
expansion can faithfully reproduce the underlying cosmological model 
up to one percent level.
Incidentally, this is the redshift range for which the large majority of
detection will be made if sources follow the realistic distribution (\ref{eq:EMobs}). Besides the parameter uncertainty $\sigma_{\theta_i}$, defined by the 68\% confidence
interval of its probability distribution function, we also report the relative bias,
defined as the absolute value of the difference between the average value recovered and injected one normalized by $\sigma_{\theta_i}$.

\begin{figure}[t]
  \centering
  \includegraphics[width=.43\linewidth]{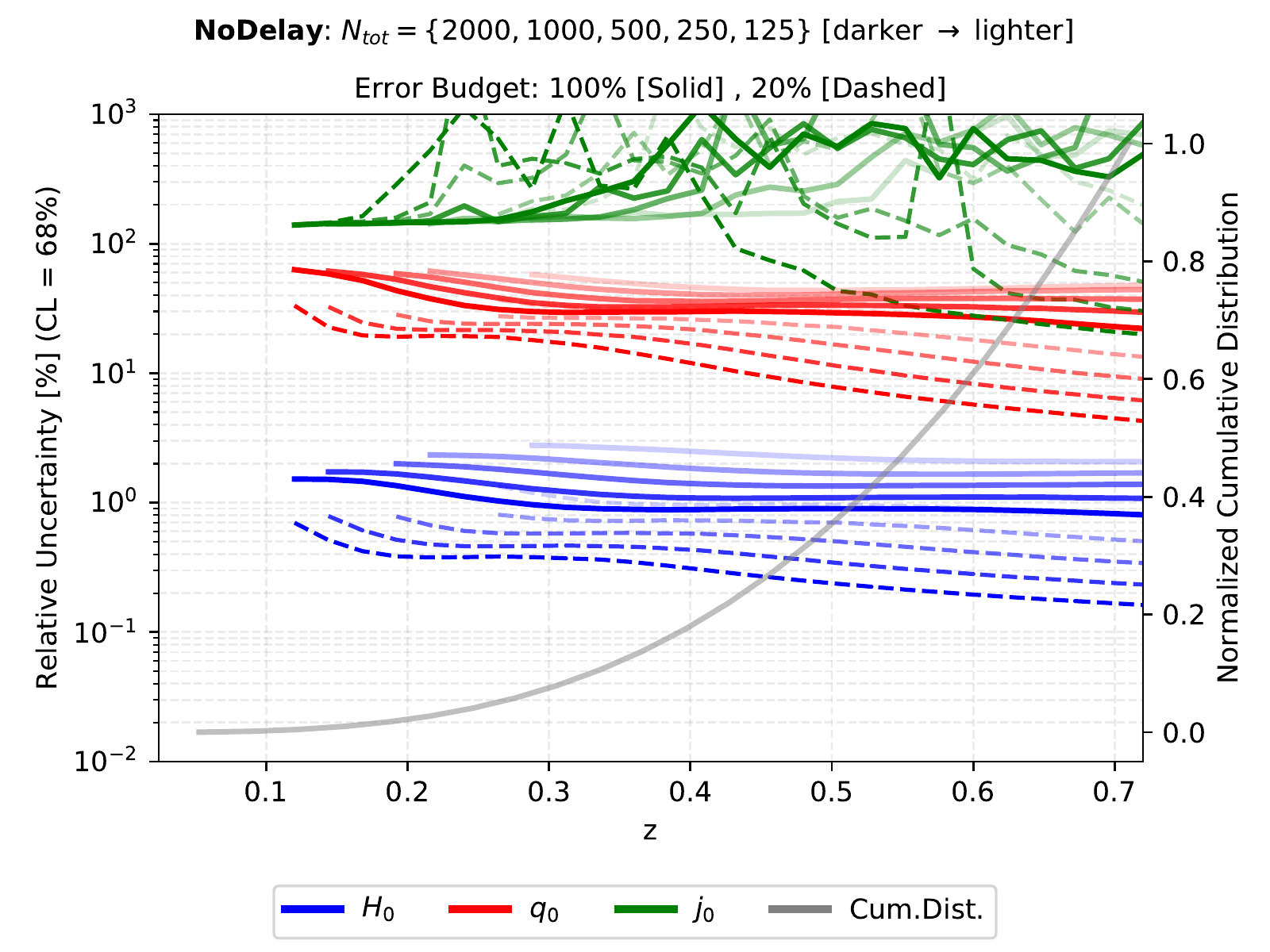}
  \includegraphics[width=.43\linewidth]{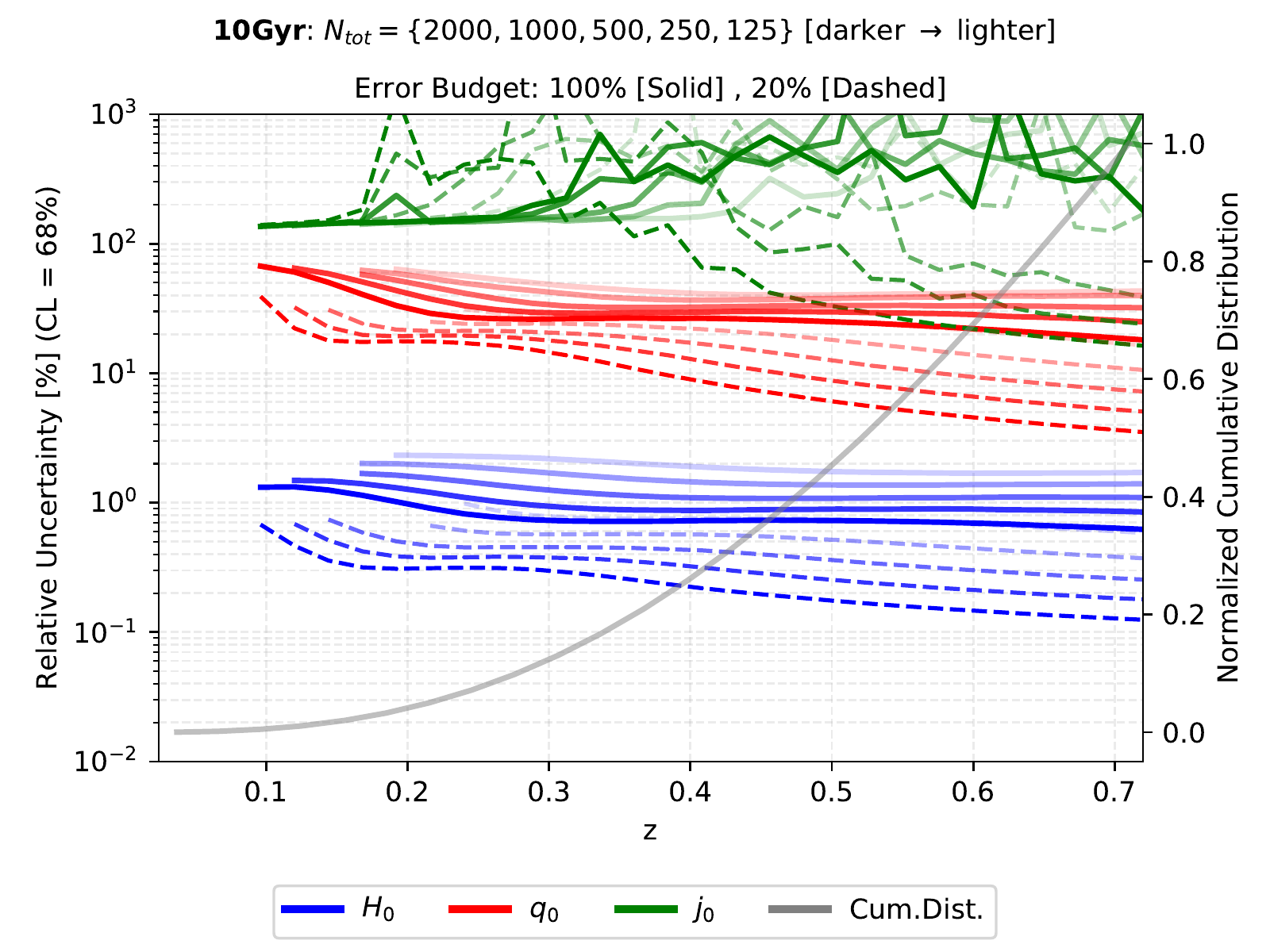}
  \includegraphics[width=.43\linewidth]{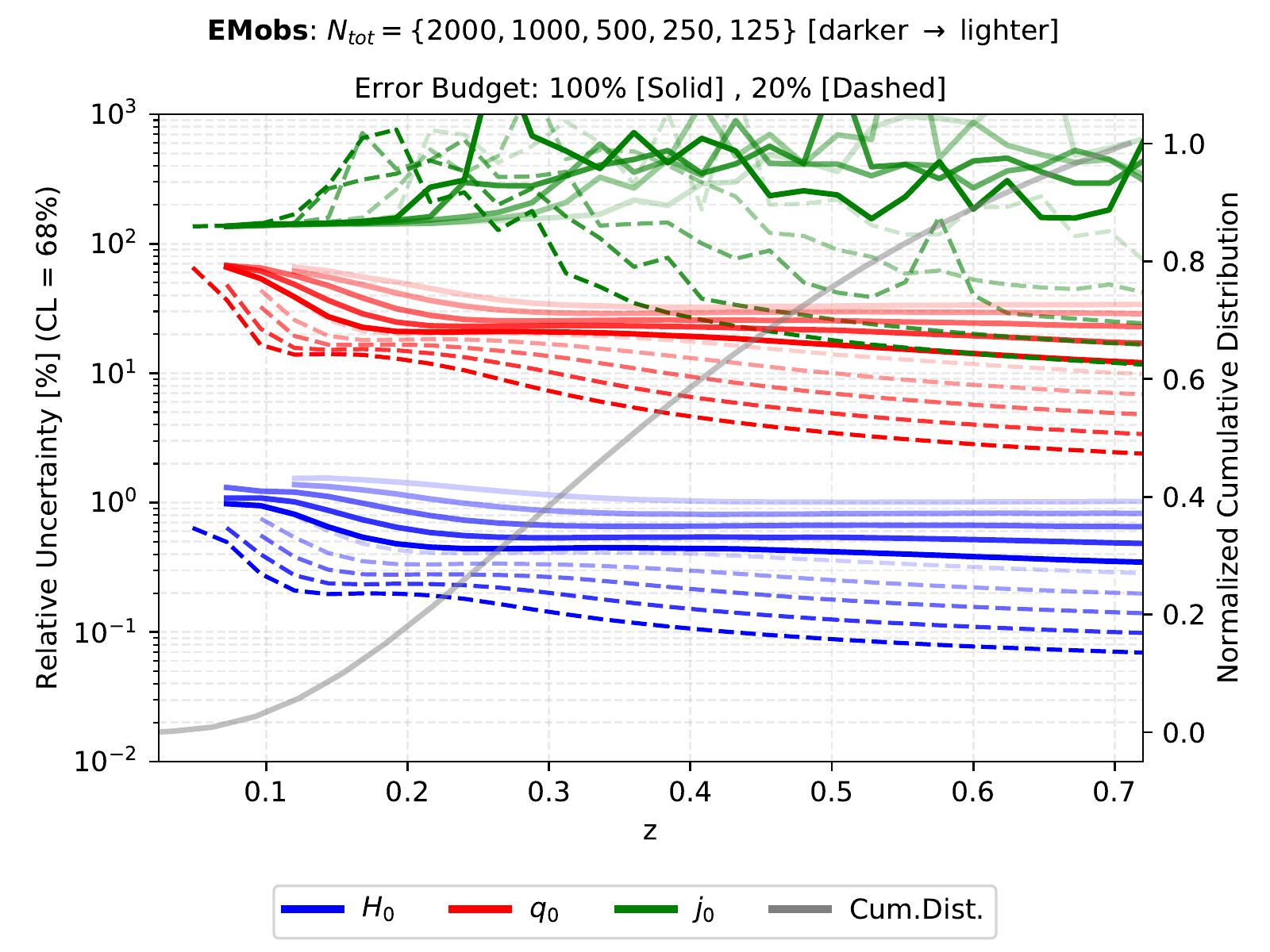}
  \caption{Results for parameter estimation error for (top) the source distribution following star formation rate (\ref{eq:sfr}), (middle) $10$ Gyr average
    delay between star formation and binary merger, and (bottom)
    EMobs (\ref{eq:EMobs}) as a function of the highest redshift detection.
    Lines get thicker as the number of total detections increases. Solid lines are for luminosity distance error as in
    Eq.~(\ref{eq:dL_err}), dot-dashed for 20\% of such value. The right vertical scale refers to the normalized cumulative distribution of detections, displayed in light grey.}
  \label{fig:result_err}
\end{figure}

\begin{figure}[t]
  \centering
  \includegraphics[width=.43\linewidth]{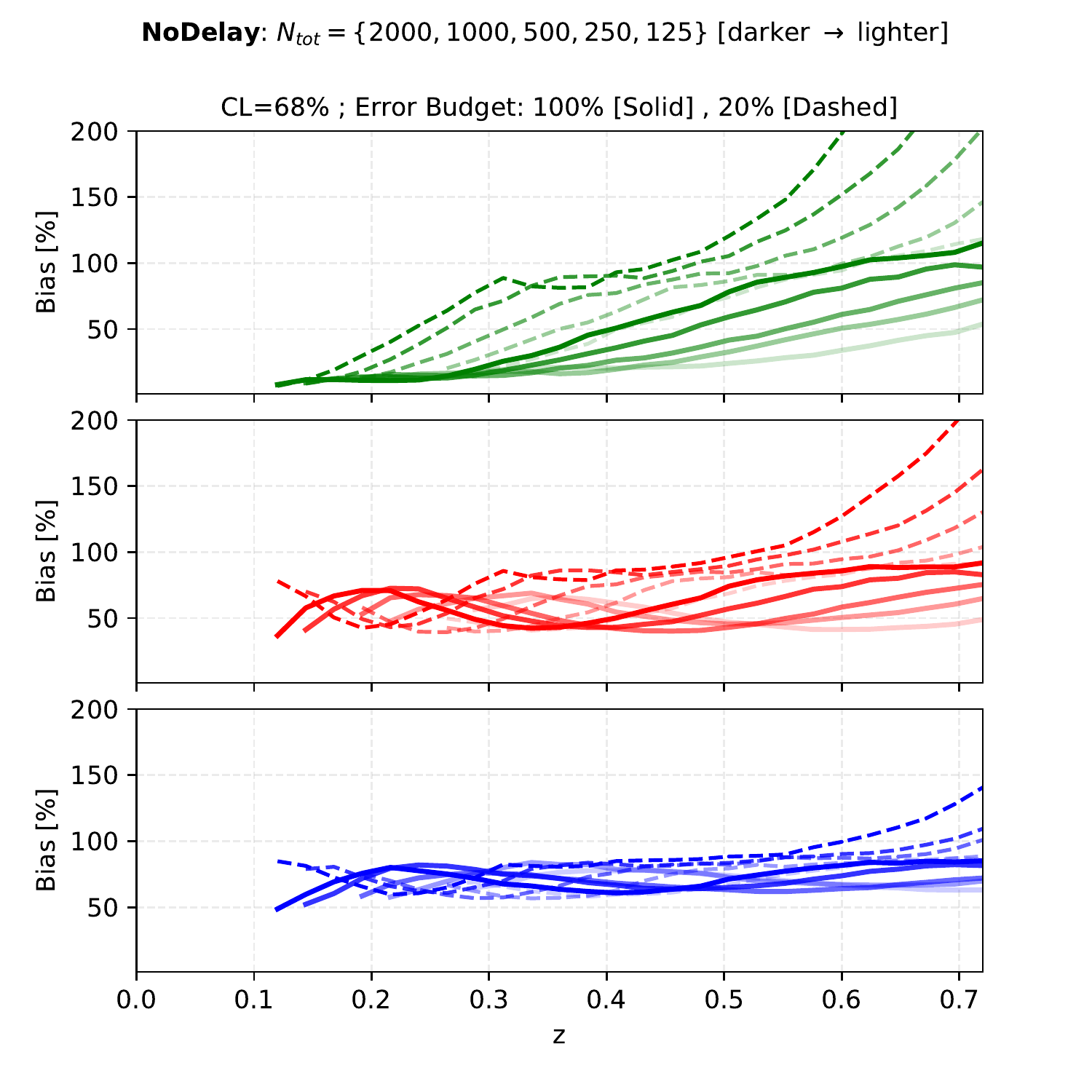}
  \includegraphics[width=.43\linewidth]{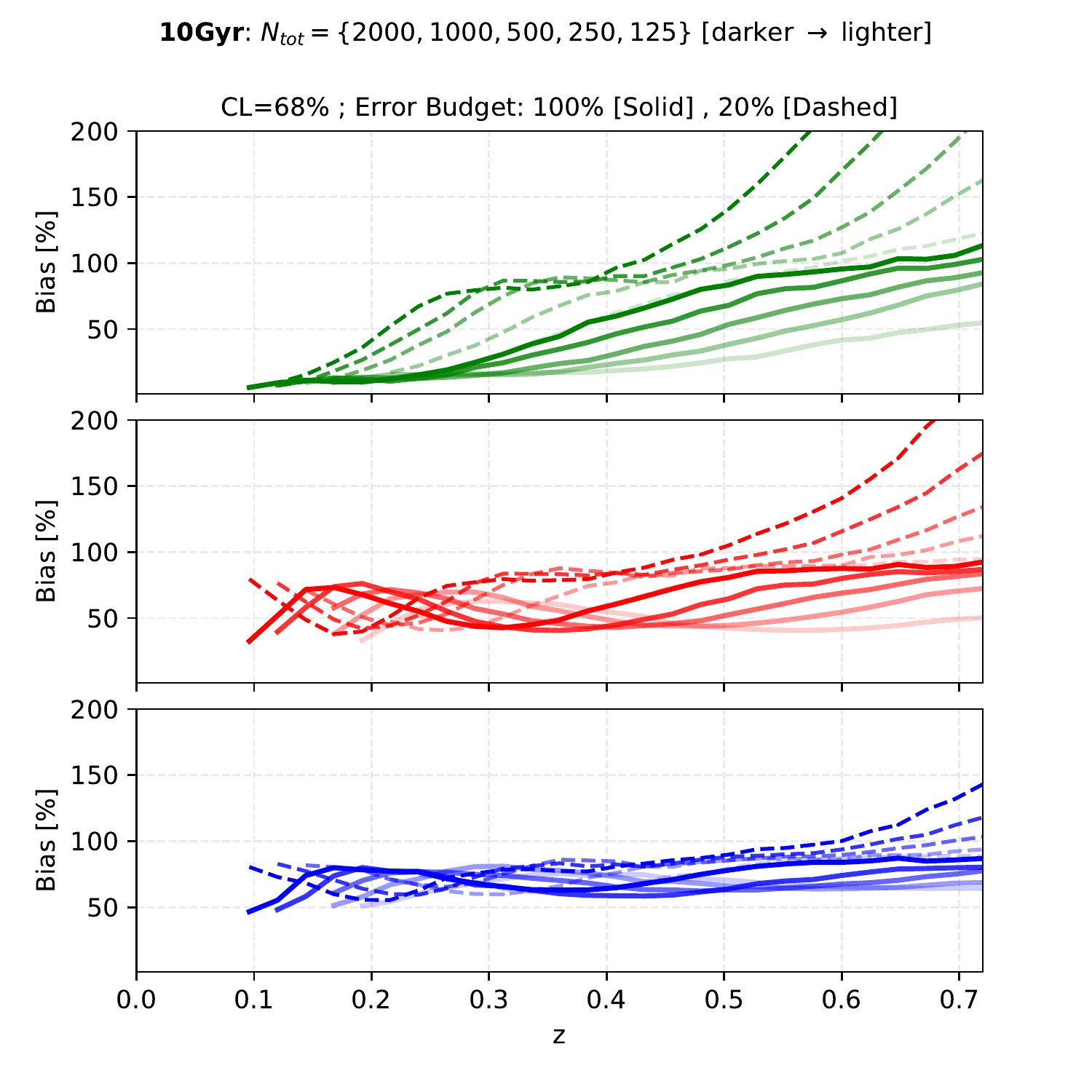}
  \includegraphics[width=.43\linewidth]{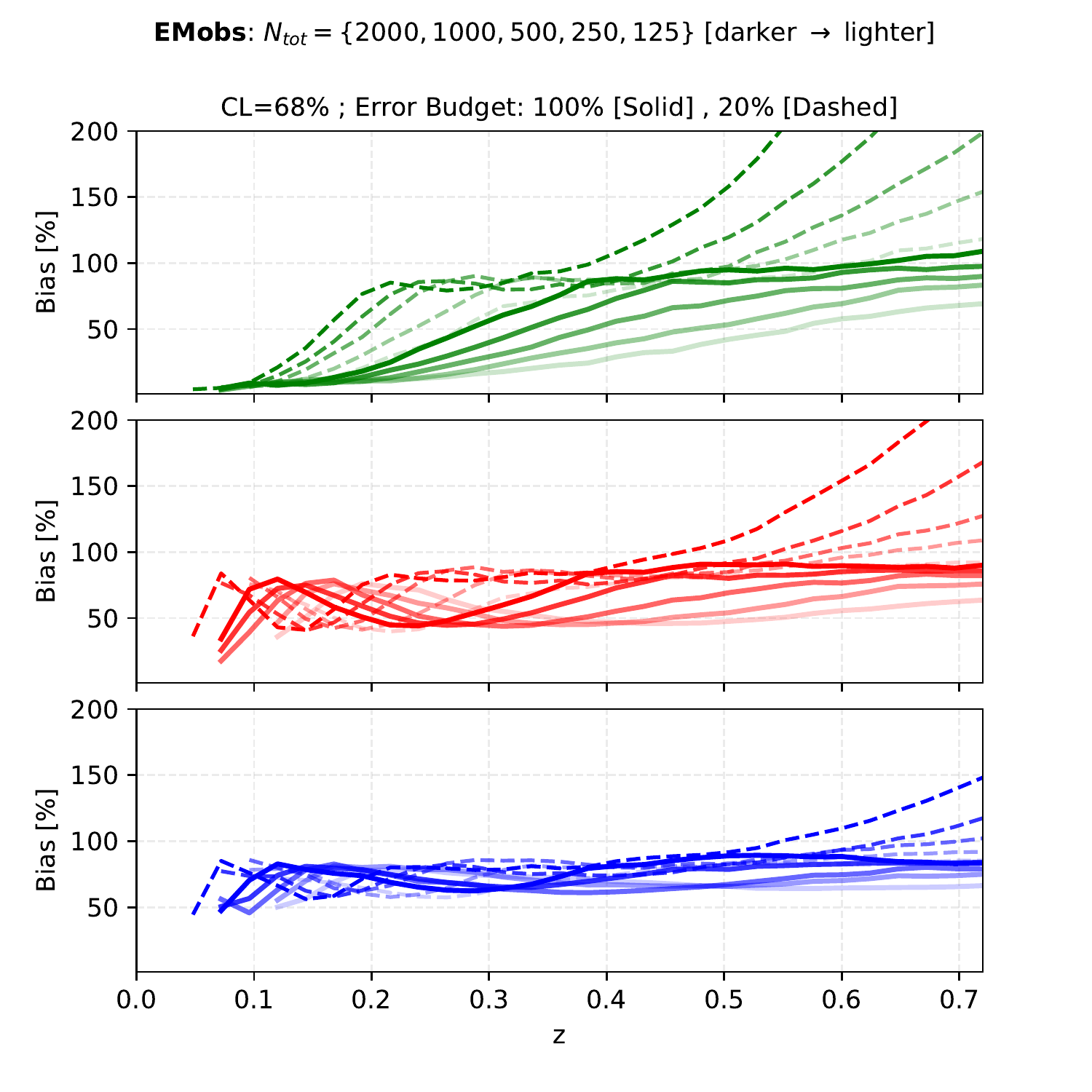}
  \caption{The same as in Fig.~(\ref{fig:result_err}) for the parameter biases.}
  \label{fig:result_bias}
\end{figure}

In Fig.~(\ref{fig:result_err}) we report the measurement of the uncertainty for the three cosmographic parameters. For each of them
we show how such error varies with the redshift range of the detections,
for five fixed values of total number of detections, ranging from 125 up to 2,000.
The total number of detections refer to the complete redshift range up to $z=0.72$,
hence the process of accumulating detections in a realistic case will
move the parameter uncertainty from lighter lines (smaller numbers
of total detections) to darker lines (larger numbers).
For completeness, we also consider two possible values of the luminosity distance uncertainties in the
measure, the value given in Eq.~(\ref{eq:dL_err}) and 20\% of that, which saturates
the lensing contribution in (\ref{eq:dL_errs}). For each set of simulations, the result is obtained by averaging over $500$
random realizations of injections extracted from the relevant redshift
distribution. Similarly, in Fig.~(\ref{fig:result_bias}) we show the bias for the three cosmographic parameters,
computed as the absolute value of the difference between the mean value and the injected value, normalized
by $\sigma_{\theta_i}$.
For each of them we report how they vary with redshift range for the same five fixed values of total number of detections.

\begin{figure}[t]
  \centering
  \includegraphics[width=.43\linewidth]{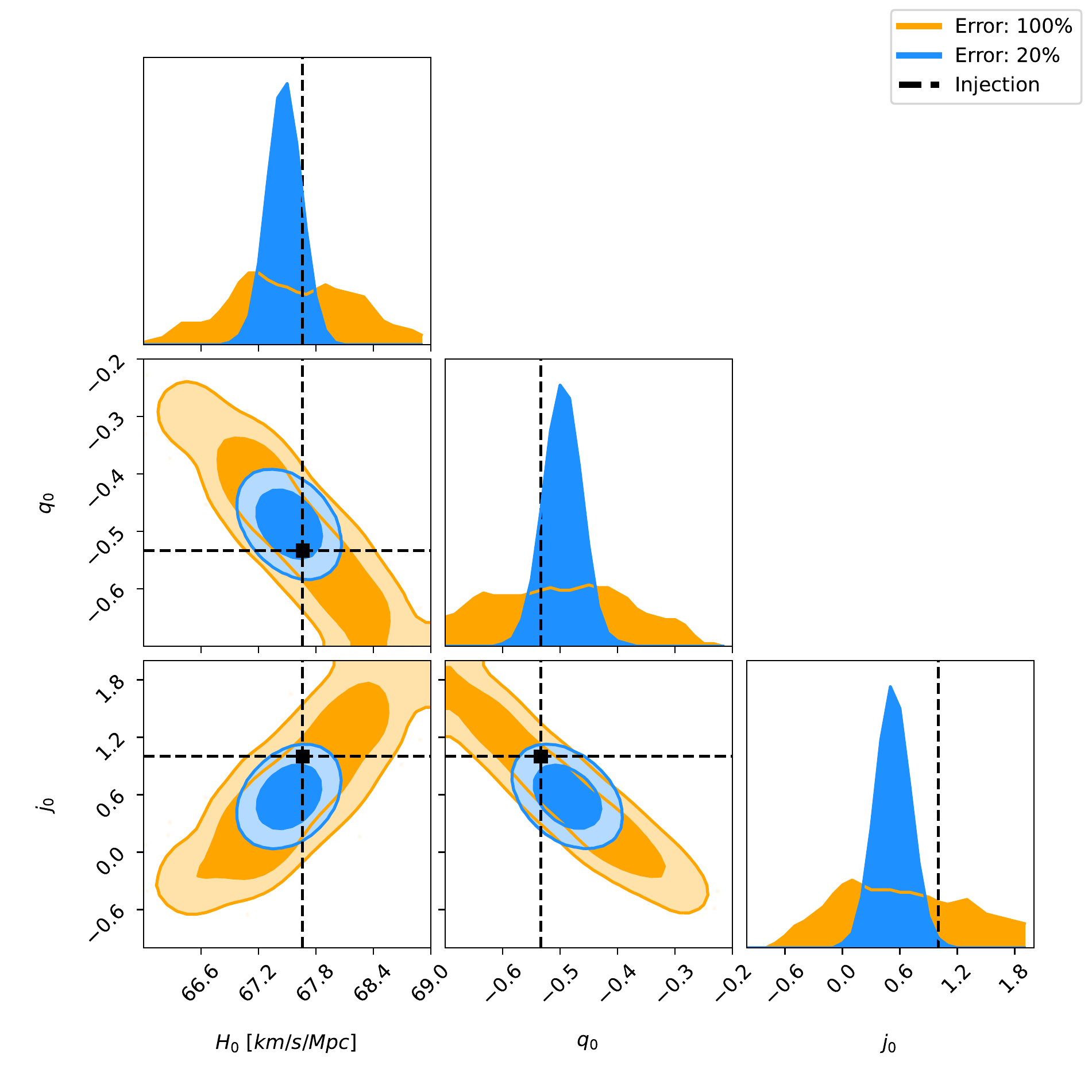}
    \includegraphics[width=.43\linewidth]{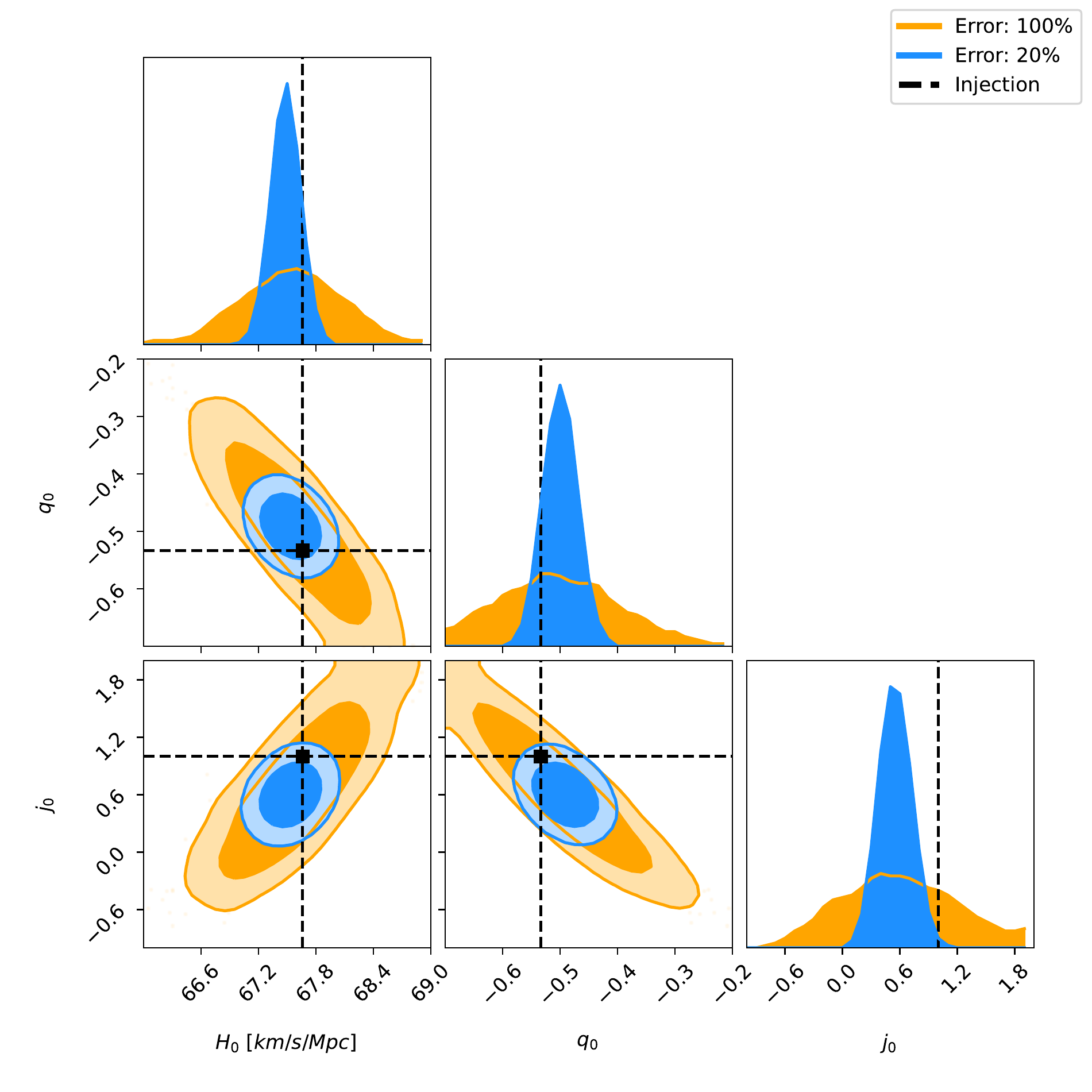}
    \includegraphics[width=.43\linewidth]{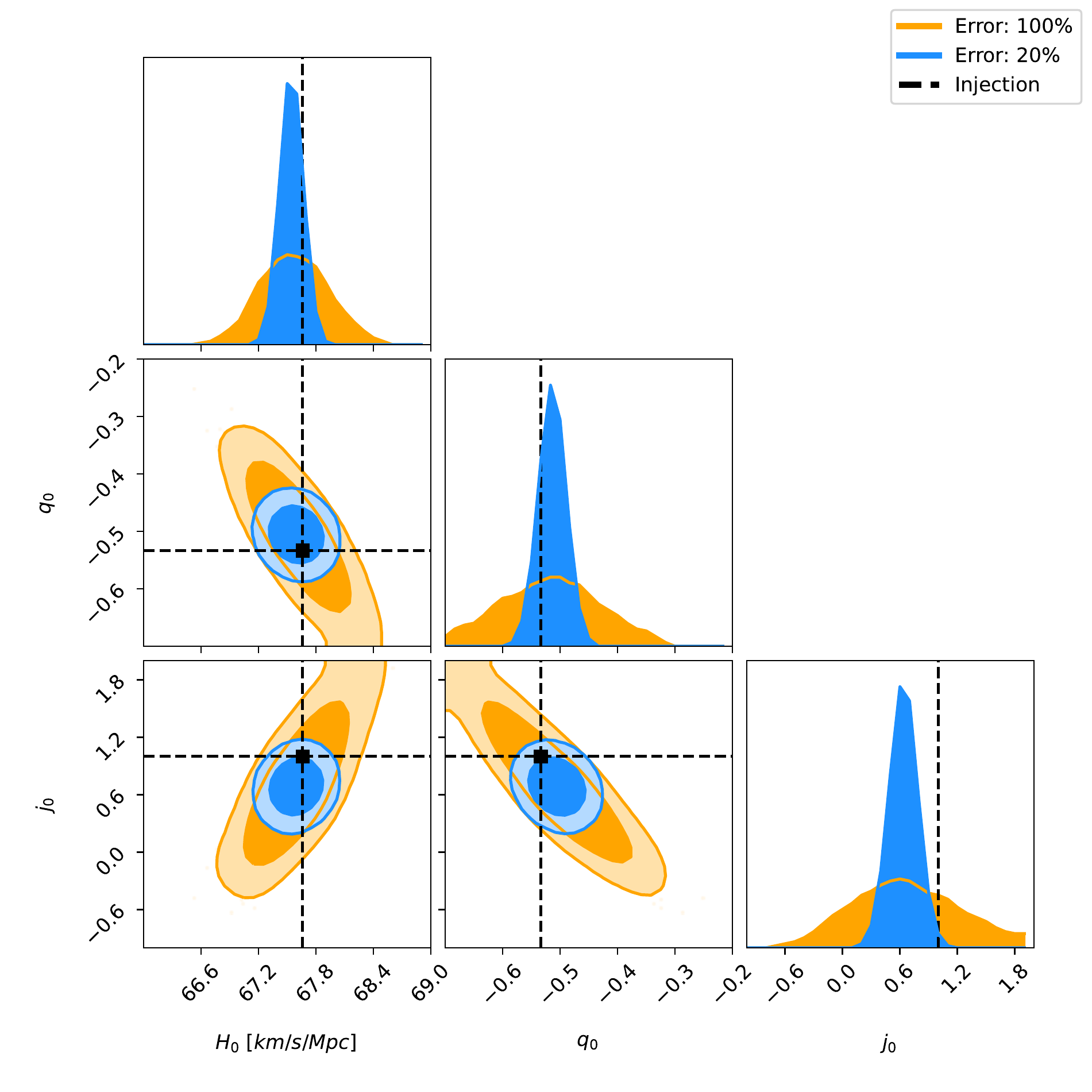}
  \caption{Marginalized posterior distributions  of the cosmographic parameter mean values for 500 realizations of 1,000 events
    drawn from No-Delay distributions (top left), $\tau=10Gyr$ delay (top right) and EMobs (bottom), see Subsec.~\ref{ssec:s_dists}. For each
  distribution the case for 100\% and 20\% the error given in Eq.~(\ref{eq:dL_errs}) are reported.}
  \label{fig:cosmo_1000_0.7}
\end{figure}

\begin{figure}[th]
    \centering
    \includegraphics[width=.43\linewidth]{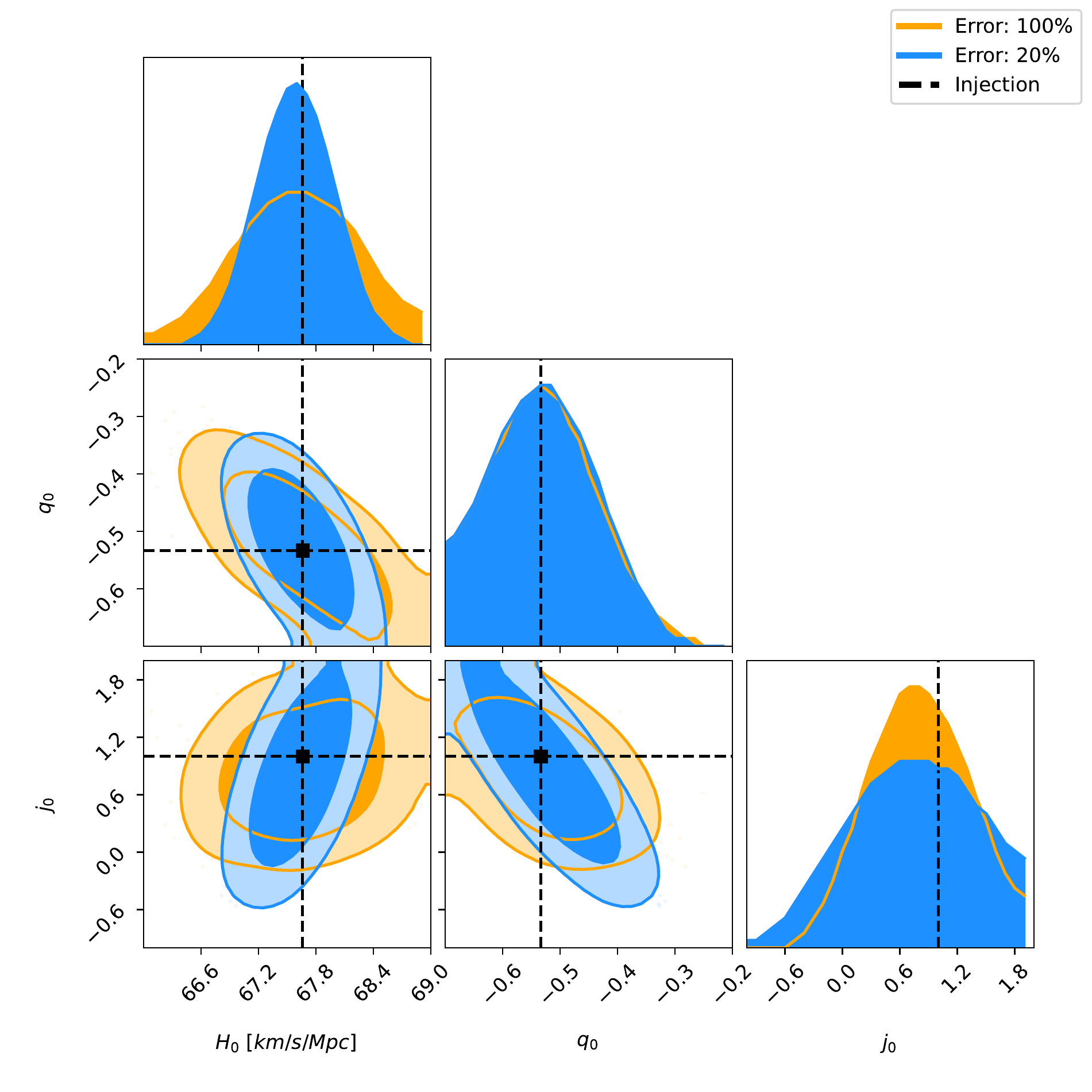}
    \includegraphics[width=.43\linewidth]{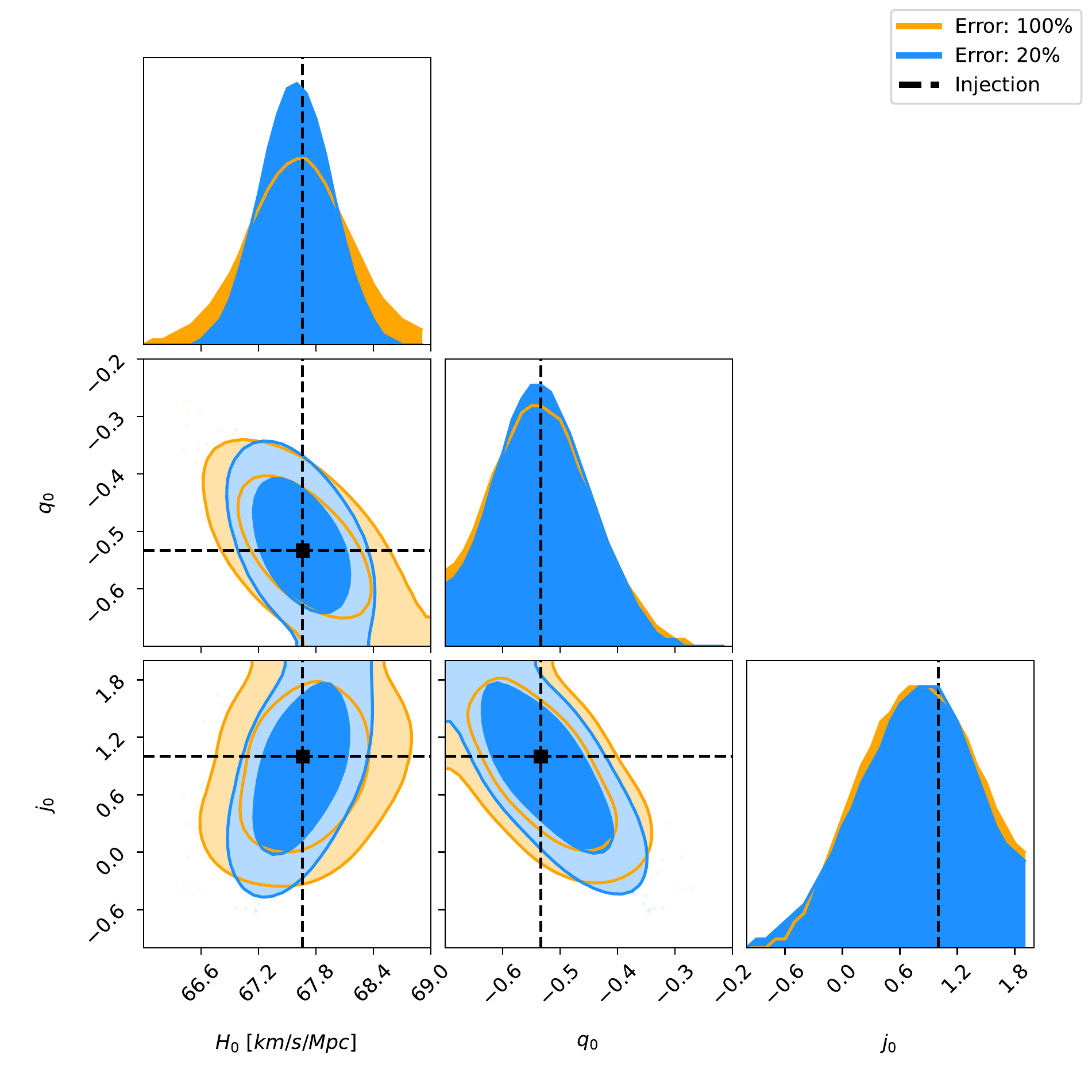}
    \includegraphics[width=.43\linewidth]{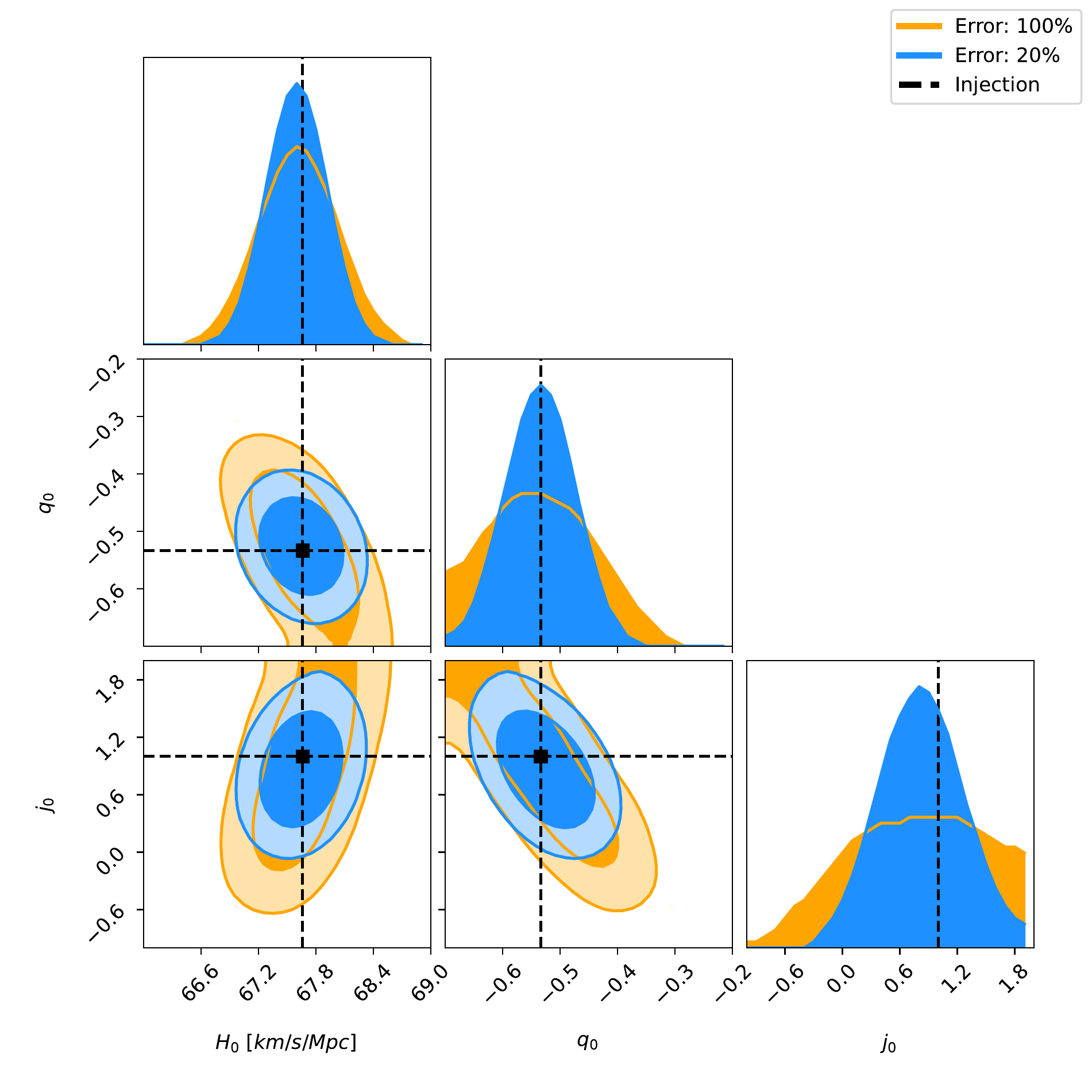}
    \caption{Same as Fig.~(\ref{fig:cosmo_1000_0.7}) for detections up to $z=0.4$,
      corresponding to $\sim $ 180, 261, 586 for no-delay, 10 Gyr and EMobs distributions respectively.}
    \label{fig:cosmo_xx_0.4}
\end{figure}

We note from Fig.~\ref{fig:result_bias} that the best fit values of the
cosmographic parameters deviates from the values of Tab.~\ref{tab:cosmopars}
when detections at redshift $z\gtrsim 0.4$ are included.
To better illustrate this aspect, we report the distributions of the best fit
values of the cosmographic parameters $H_0$, $q_0$, and $j_0$
for the different source distributions discussed in  Subsec.~\ref{ssec:s_dists}
obtained over 500 realizations, for detections up to redshift $z_c=0.71$ in
Fig.~\ref{fig:cosmo_1000_0.7}, and up redshift $0.4$ in Fig.~\ref{fig:cosmo_xx_0.4}.
Clearly detections at higher redshift tend to shift the best fit to values
incompatible to those obtained via the exact cosmographic expansion of
$\Lambda$CDM given in Tab.\ref{tab:cosmopars}, the effect being stronger
for astrophysical distributions having more events at high redshift.
However the bias effect is visible only when the measure error of the
luminosity distance is reduced to the lensing contribution, i.e. to 20\% its
realistic value.

\begin{figure}[t]
  \centering
  \includegraphics[width=.65\linewidth]{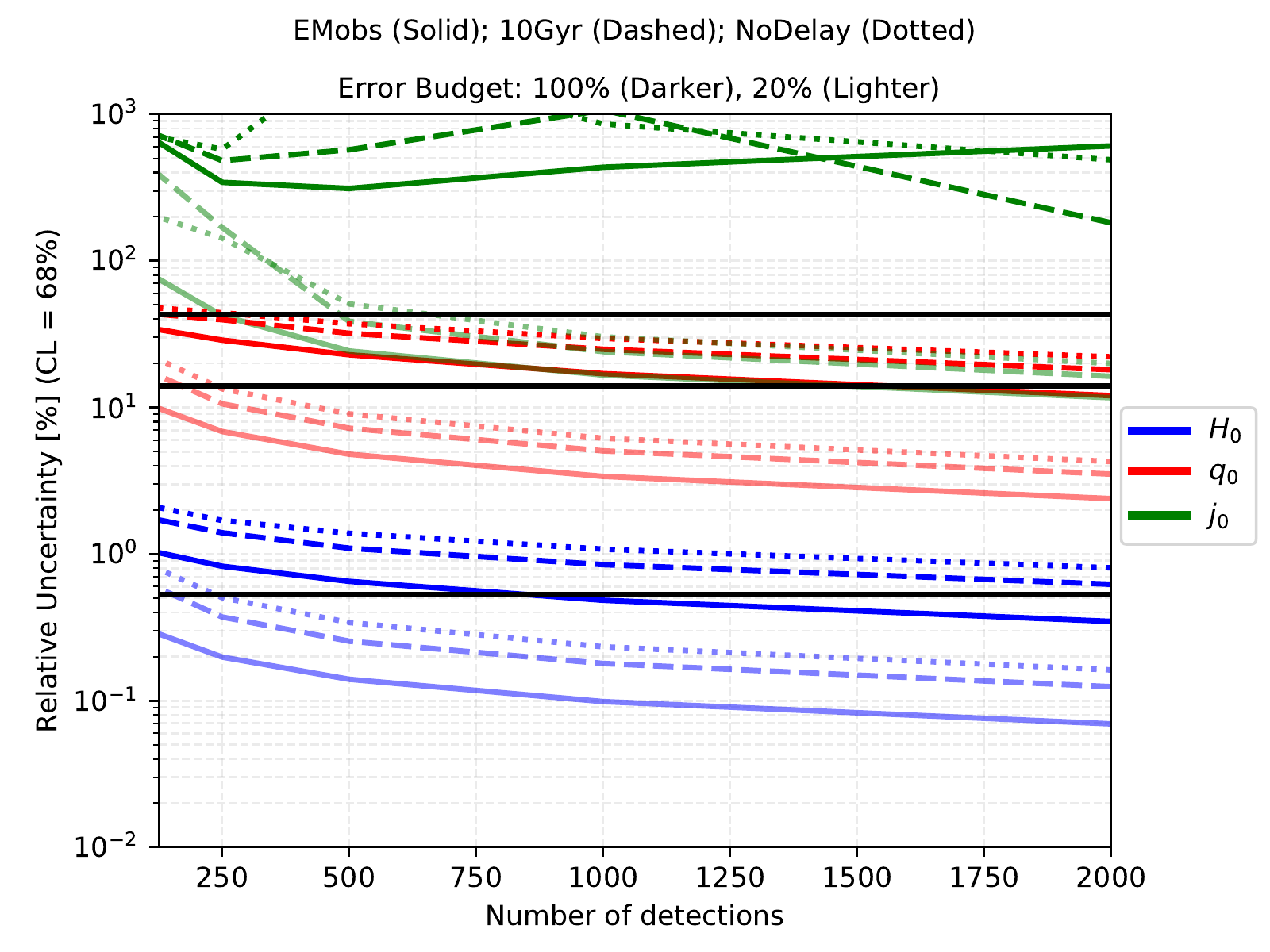}
  \caption{Comparisons between the error estimations for $H_0$, $q_0$, and $j_0$
    from standard sirens (for three source distributions and different number of detections) and SNe data \cite{2018ApJ...859..101S} (horizontal black lines from bottom to top). In all cases only SNe up to redshift  $z_{cut}=0.72$ have been considered, which correspond to 972 detections.}
  \label{fig:SN_GW}
\end{figure}

Finally, we also find useful to compare the results obtained from GW detections
with the current estimates of the first three cosmographic parameters from type Ia supernovae (SNe) data \cite{2018ApJ...857...51J}.
We adopt the same cutoff in redshift, $z_{cut}=0.72$, which results in a SNe subsample with 972 data points,
and fix the absolute peak magnitude at  $M_B = -19.214 \pm 0.037$ mag, as given by \cite{Riess:2020fzl}.
Our analysis gives $H_0=75.23^{+0.40}_{-0.38}$ km/sec/Mpc, $q_0=-0.70 \pm 0.10$, $j_0=1.78^{+0.78}_{-0.74}$,
with the error bars corresponding to 68\% confidence limit\footnote{Note that similar uncertainties on $H_0$ were obtained in \cite{Chang:2019xcb} for a comparable number of detections within the $\Lambda$CDM cosmology rather than the 
  cosmographic modeling.}.
As shown in Fig.~(\ref{fig:SN_GW}),  GWs detections can reach a better precision than current SNe data in measurements of $H_0$ for a number of detection larger
than about 1,000 considering the standard errors on $d_L$, and for $\sim$ 100 detections when the errors are reduced to 20\%, the value
of Eq.~(\ref{eq:dL_errs}), see also Tab.~\ref{tab:dist_Ns100} and Tab.~\ref{tab:dist_Ns20}.

\begin{table}[t]
	\centering
\begin{tabular}{|c|cc|cc|cc|}
		\hline
	  {} & \multicolumn{2}{|c|}{$|\Delta H_0|$ km/s/Mpc} & \multicolumn{2}{|c|}{$|\Delta q_0|$} & \multicolumn{2}{|c|}{$|\Delta j_0|$} \\ \hline
	  {} & 500 sirens & 1000 sirens & 500 sirens & 1000 sirens & 500 sirens & 1000 sirens \\ \hline
	 EMobs 	 & 0.44  &  0.33  &  0.11  &  0.08  &  0.74  &  0.55 \\
	 10Gyr 	 & 0.75  &  0.58  &  0.15  &  0.12  &  0.84  &  0.66 \\
	 NoDelay & 0.95  &  0.74  &  0.17  &  0.14  &  0.93  &  0.74 \\ \hline
\end{tabular}
\caption{$1$-$\sigma$ parameter uncertainties $\sigma_{\theta_i}$ from 500 and 1000 standard sirens detections up to redshift $z=0.72$, for $d_L$ measurement
  uncertainty given by Eq.~(\ref{eq:dL_err})
  for sources distributed according to the \textit{EMobs}, \textit{10Gyr} and \textit{NoDelay} distributions.}\label{tab:dist_Ns100}
\end{table}

\begin{table}[t]
	\centering
\begin{tabular}{|c|cc|cc|cc|}
\hline
	  {} & \multicolumn{2}{|c|}{$|\Delta H_0|$ (km/s/Mpc)} & \multicolumn{2}{|c|}{$|\Delta q_0|$} & \multicolumn{2}{|c|}{$|\Delta j_0|$} \\ \hline
	  {} & 500 sirens & 1000 sirens & 500 sirens & 1000 sirens & 500 sirens & 1000 sirens \\ \hline
	 EMobs 	 & 0.09  &  0.07  &  0.02  &  0.02  &  0.16  &  0.11 \\
	 10Gyr 	 & 0.17  &  0.12  &  0.04  &  0.02  &  0.20  &  0.14 \\ 
	 NoDelay & 0.23  &  0.16  &  0.04  &  0.03  &  0.23  &  0.16 \\ \hline
\end{tabular}
\caption{$1$-$\sigma$ parameter uncertainties $\sigma_{\theta_i}$ from 500 and 1000 standard sirens detections up to redshift $z=0.72$ with $d_L$ measurement
  uncertainty reduced to 20\%, for sources distributed according to
  \textit{EMobs}, \textit{10Gyr} and \textit{NoDelay} distributions.}
\label{tab:dist_Ns20}
\end{table}

\section{Conclusions}
\label{sec:concl}
We analyzed how future detections of gravitational wave standard sirens by a third generation detector
can measure the Hubble constant, the deceleration and the jerk parameters of the cosmographic expansion.
As it is well known, truncated cosmographic expansions are not supposed to describe correctly the cosmological
expansions beyond redshift of order unity,
and we showed that our choice for $z_{cut}=0.72$ leads to a faithful cosmographic model with only the first
three cosmographic parameters.
The simulated data assumed the contemporary observation of electromagnetic counterparts ensuring redshift determination,
and overall detections up to moderate redshift $z_{cut}$.

The results show that the Hubble constant can be measured with sub-percent precision with a few hundred of detections.
For higher order parameter in the cosmographic expansion precision is degraded, e.g. for $q_0$ one can reach
10\% level with 2,000 detections, whereas for $j_0$ it is hard to limit
precision below 100\% level.

Our analysis showed an important caveat, i.e. the best fit values
  of the third order cosmographic expansion tend to drift away from the exact
  cosmographic parameter values obtained by a complete Taylor expansion of
  the underlying $\Lambda$CDM model, when detections at redshit $z\gtrsim 0.4$
  are included.
  While this effect is contained within the 1-$\sigma$ prediction errors for
  realistic predictions of the luminosity distance measure uncertainty,
  it is visible when one uses a reduced measurement error on $d_L$, corresponding
  e.g. to the uncertainty induced by lensing of matter structure between
  source and observer.
  The reference cosmographic parameters, summarized in Tab.~\ref{tab:cosmopars},
  are obtained using the full Taylor expansion of the $d_L$ vs. $z$ relation
  in the $\Lambda$CDM model, hence it is not surprising that the best fit values
  we obtain for the three cosmographic parameters deviates from them, as
  also illustrated by Figs.\ref{fig:RelDiff100},\ref{fig:RelDiff20}
  in Appendix \ref{app:stab}.

We also analyzed how such results vary with the distribution of events,
showing that a distribution more concentrated at low redshifts allows a sharper determination of the $H_0$ and $q_0$.
Besides, the error in the parameter determination is mostly determined by the detections at low redshifts ($\lesssim 0.2$),
and then levels off accumulating detections at higher redshifts unless thousands of events are accumulated.
We also varied the number of expected detections, whose rate at present is wildly unknown,
between realistic values of $O(10^2)$ and $O(10^3)$, showing an expected monotonic,
sharpening of precision with increasing number of detection.
Our projection for the $H_0$ measure precision with $\sim$ 1,000 sources is of the same order the one
that can be obtained analyzing the presently available SNe data up to $z_{cut}$, as shown in Fig.~\ref{fig:SN_GW}, its exact value varying non-negligibly with the number of detections,
their redshift distributions and luminosity distance measurement uncertainty.

Finally, we showed that a crucial ingredient to reduce the measurement error in the cosmographic parameters is represented by the
observational uncertainty in $d_L$, which can in principle be reduced by
correlating observations
from separately located detectors, to reach the ``lensing limit'', which at
moderate redshifts amounts to about 20\% of the luminosity distance error budget
for a single interferometer.

\appendix
\section{${\mathbf \Lambda}$CDM versus cosmographic expansion }
\label{app:stab}

\begin{figure}[t]
  \begin{center}
    \includegraphics[width=.49\linewidth]{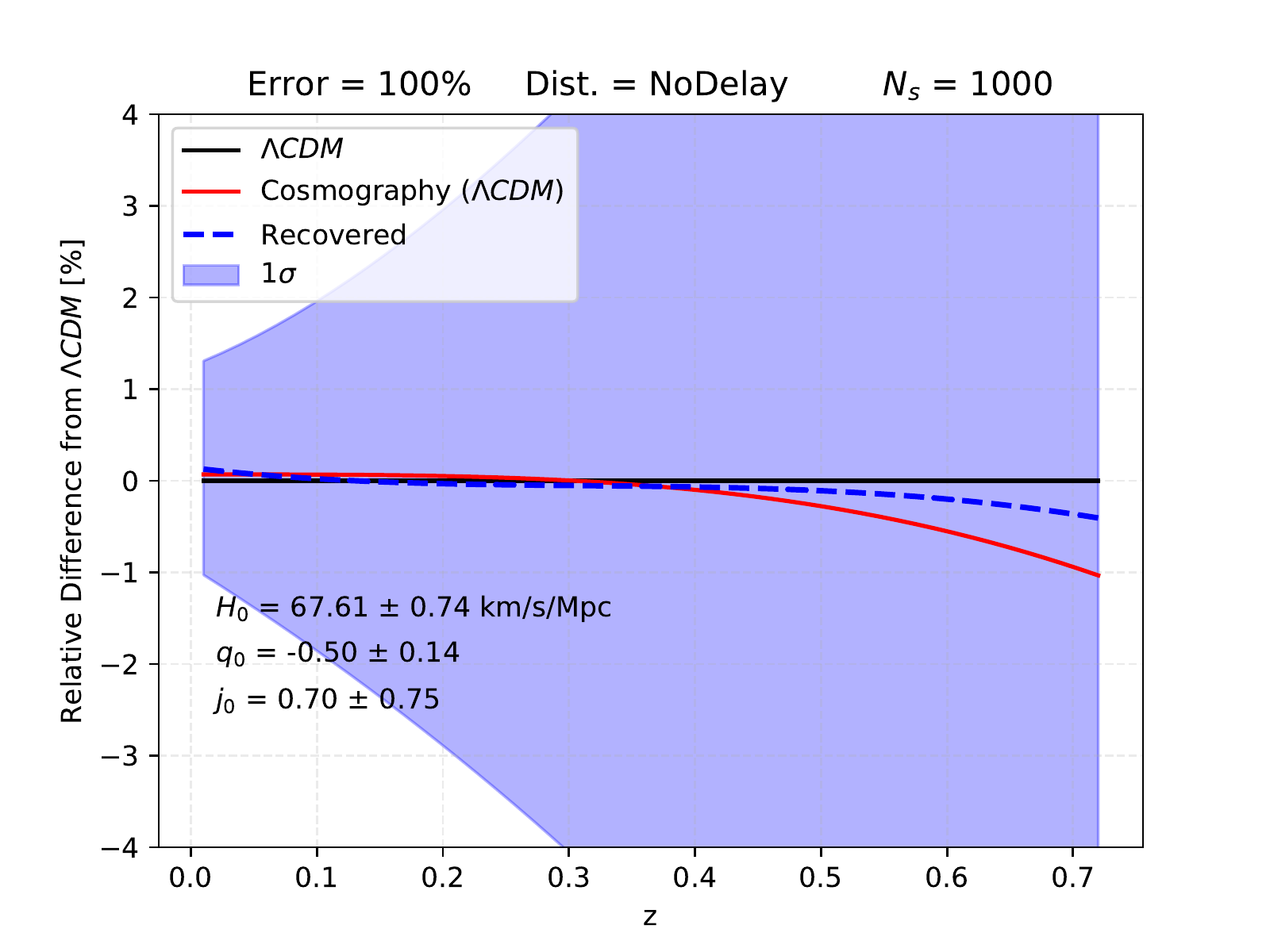}
    \includegraphics[width=.49\linewidth]{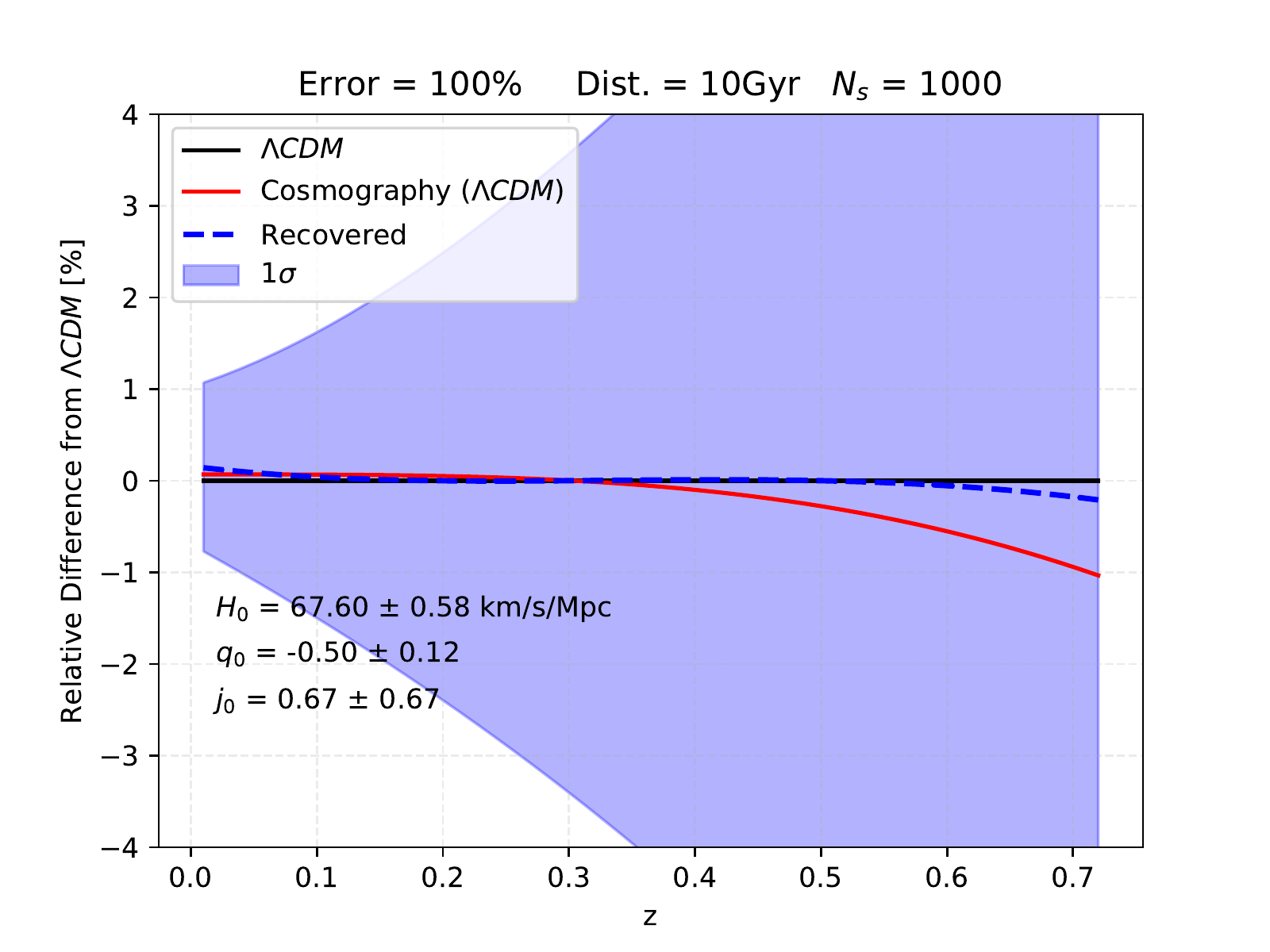}
    \includegraphics[width=.49\linewidth]{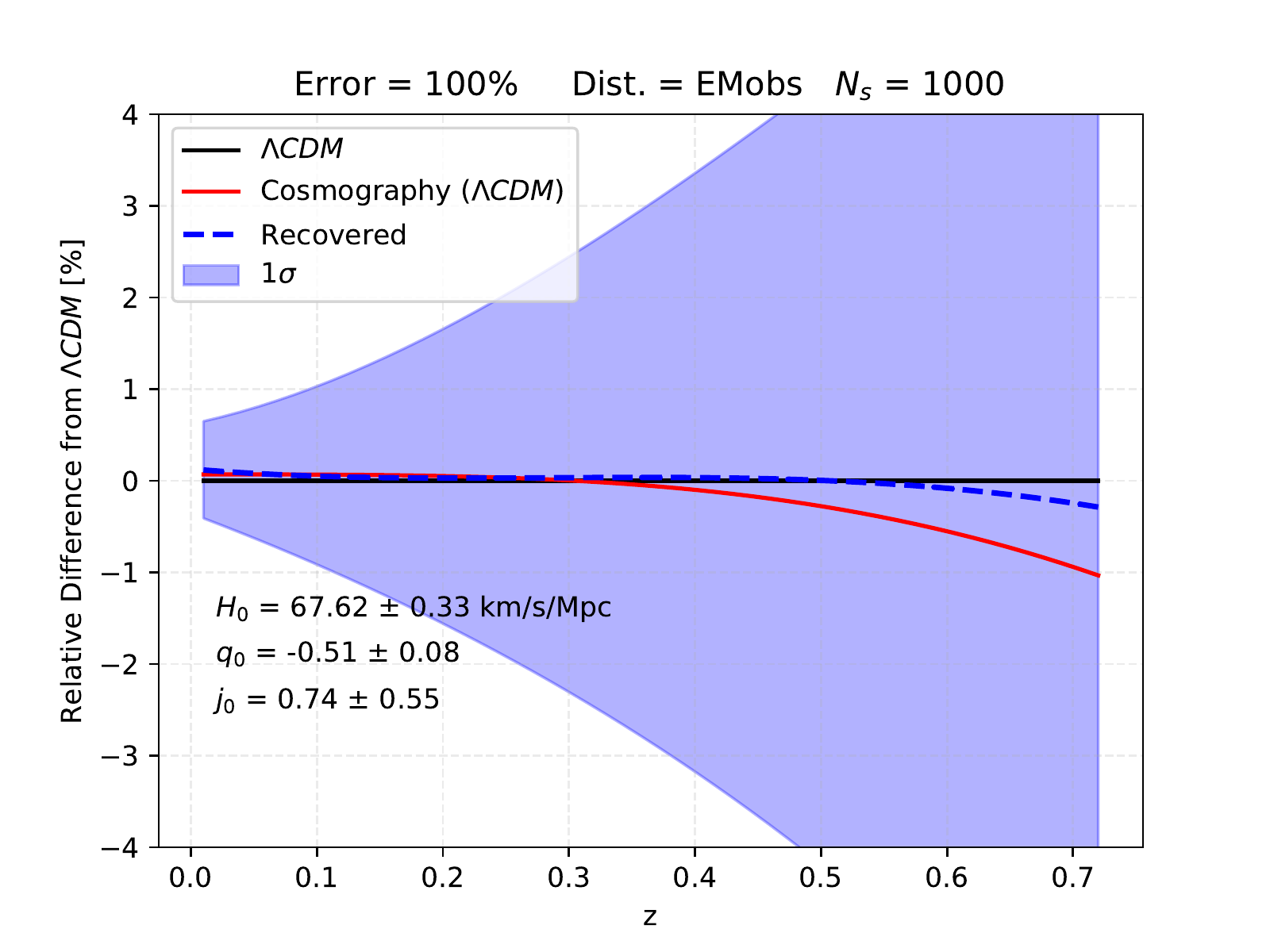}
    \caption{Relative difference between the luminosity distance according to
      the fiducial model (flat $\Lambda CDM$) and the one recovered from the
      Bayesian inference with the cosmographic parameters given in Tab.~\ref{tab:dist_Ns100} (blue dashed).
      Also reported is the difference between the fiducial model and
      the cosmographic one with values of the three cosmographic
    parameters taken from Tab.~\ref{tab:cosmopars} (red).}
    \label{fig:RelDiff100}
  \end{center}
\end{figure}

\begin{figure}[t]
  \begin{center}
  \includegraphics[width=.49\linewidth]{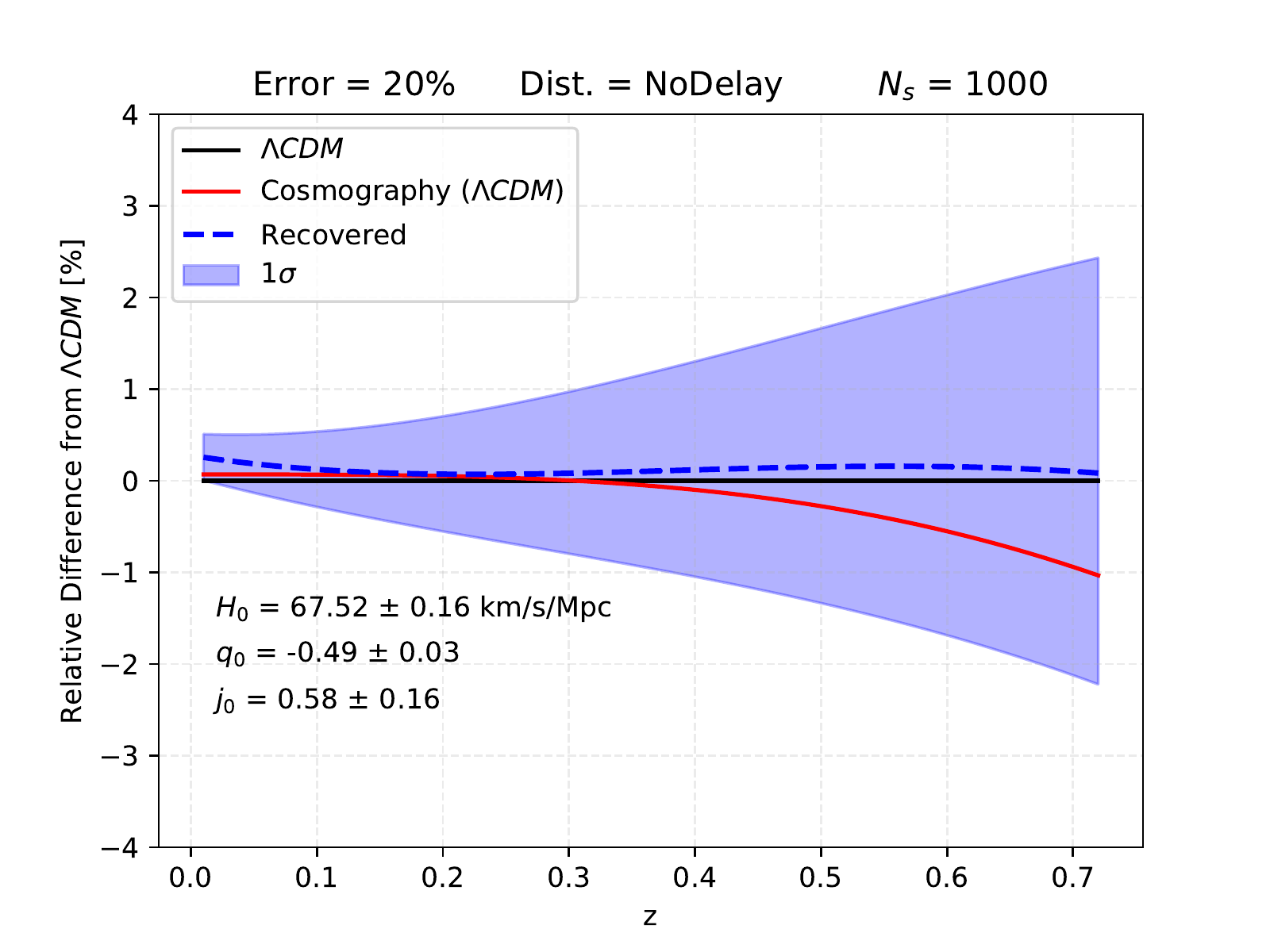}
  \includegraphics[width=.49\linewidth]{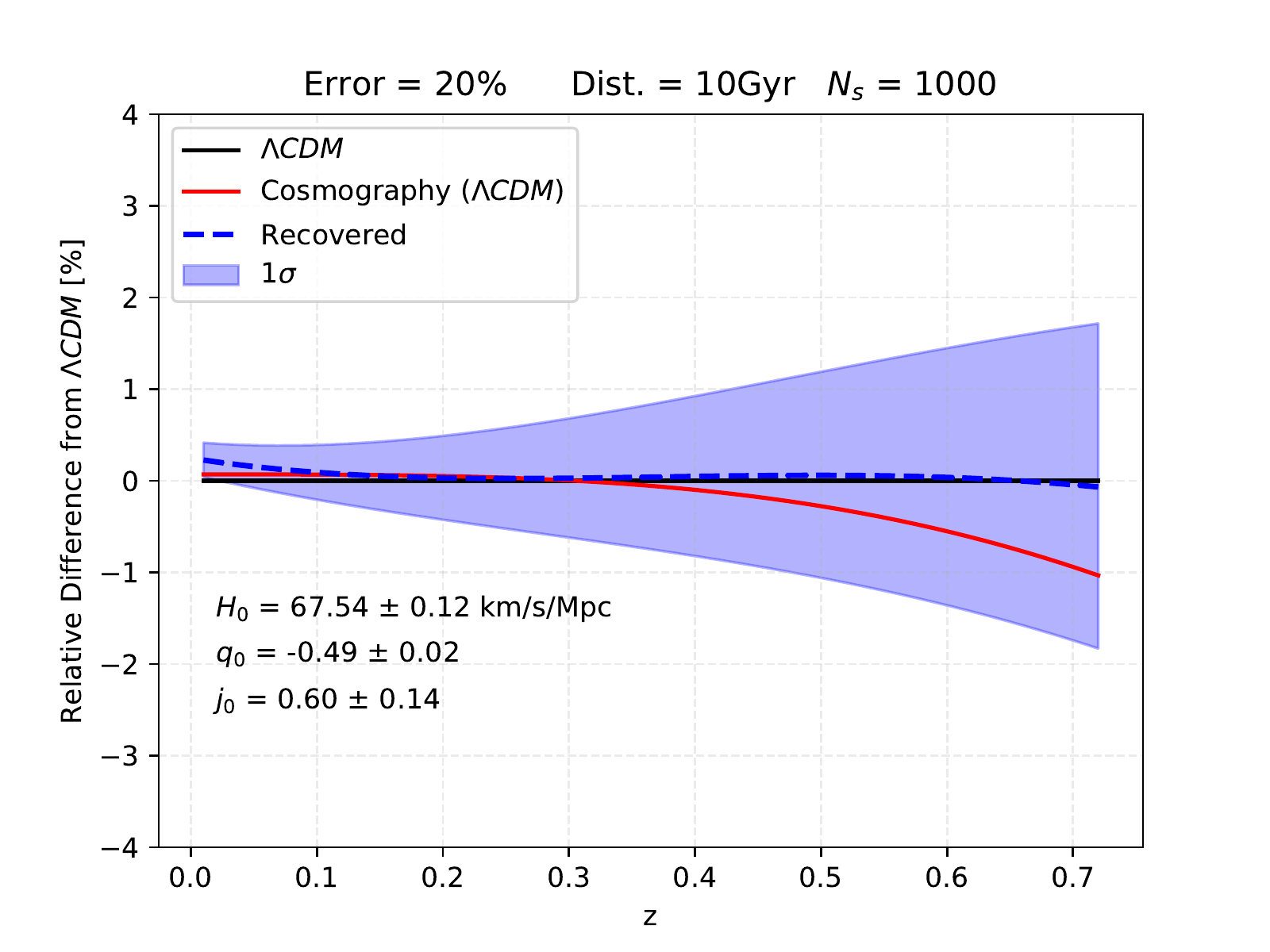}
  \includegraphics[width=.49\linewidth]{Figures/20_10Gyr_1000.pdf}
  \caption{Same as Fig.~\ref{fig:RelDiff100} from simulated data with
    reduced $d_L$ uncertainty measure.}
  \label{fig:RelDiff20}
  \end{center}
\end{figure}

In Figs.~\ref{fig:RelDiff100}-\ref{fig:RelDiff20} we show the relative
difference between the luminosity distance from the fiducial $\Lambda$CDM
model and from the best fit cosmographic one (dashed blue line)
and its $1$-$\sigma$ uncertainty.
For comparison we also report the differernce between the $\Lambda$CDM model
and a cosmographic third order model using the values of $H_0$, $q_0$ and $j_0$
given in Tab. \ref{tab:cosmopars}.
Figs.~\ref{fig:RelDiff100} refer to the three source redshift distributions
listed in Subsec.~\ref{ssec:s_dists} for realistic luminosity distance
uncertainties, and Figs.~\ref{fig:RelDiff100} to the case of reduced $d_L$
measurement uncertainty.

%---------------------------------------------
%---------------------------------------------

\section*{Acknowledgements}
JMSdS is supported  by the Coordenação de Aperfeiçoamento de Pessoal de Nível Superior (CAPES)  -- Graduate Research Fellowship/Code 001. The work of RS is partly supported by CNPq under grant 312320/2018-3. JSA acknowledges support from CNPq  No.~310790/2014-0 and FAPERJ No.~E-26.200.842/2021. The authors  thank  the  High  Performance  Computing  Center  (NPAD)  at  UFRN for providing computational resources.

\providecommand{\href}[2]{#2}\begingroup\raggedright\endgroup

\end{document}